\documentclass[aps,prb,twocolumn,twoside,superscriptaddress,longbibliography]{revtex4-1}

\usepackage[version=3]{mhchem} 
\usepackage{graphicx}
\usepackage{amsmath}
\usepackage{float}
\usepackage{amssymb}
\usepackage{color}
\usepackage{changes}
\usepackage{mathrsfs}
\usepackage{comment}

\newcommand{\sub}[1]{\ensuremath{_{\mathrm{#1}}}}
\newcommand{\super}[1]{\ensuremath{^{\mathrm{#1}}}}
\renewcommand{\vec}[1]{\ensuremath{\mathbf{#1}}}

\begin{document}

\title{\emph{Ab initio} Ultrafast Spin Dynamics in Solids}      \setcounter{page}{1}  
\date{\today}  
\author{Junqing Xu\footnote[3]{JX and AH contributed equally to this work.}} \affiliation{Department of Chemistry and Biochemistry, University of California, Santa Cruz, CA 95064, USA}      \author{Adela Habib\footnotemark[3]} \affiliation{Department of Physics, Applied Physics and Astronomy, Rensselaer Polytechnic Institute, 110 8th Street, Troy, New York 12180, USA}      \author{Ravishankar Sundararaman\footnote[2]{sundar@rpi.edu}}  \affiliation{Department of Materials Science and Engineering, Rensselaer Polytechnic Institute, 110 8th Street, Troy, New York 12180, USA}      \author{Yuan Ping\footnote[1]{yuanping@ucsc.edu}} \affiliation{Department of Chemistry and Biochemistry, University of California, Santa Cruz, CA 95064, USA} 
\begin{abstract}

Spin relaxation and decoherence is at the heart of spintronics and
spin-based quantum information science. Currently, theoretical approaches
that can accurately predict spin relaxation of general solids including
necessary scattering pathways and capable for ns to ms simulation
time are urgently needed. We present a first-principles real-time
density-matrix approach based on Lindblad dynamics to simulate ultrafast
spin dynamics for general solid-state systems. Through the complete
first-principles descriptions of pump, probe and scattering processes
including electron-phonon, electron-impurity and electron-electron
scatterings with self-consistent electronic spin-orbit
couplings, our method can directly simulate the ultrafast pump-probe
measurements for coupled spin and electron dynamics over ns at any
temperatures and doping levels. We first apply this
method to a prototypical system GaAs and obtain excellent agreement
with experiments. We found that the relative contributions of different
scattering mechanisms and phonon modes differ considerably between
spin and carrier relaxation processes. In sharp contrast to previous
work based on model Hamiltonians, we point out that the electron-electron
scattering is negligible at room temperature but becomes dominant
at low temperatures for spin relaxation in n-type GaAs. 
We further examine ultrafast dynamics in novel spin-valleytronic
materials - monolayer and bilayer WSe$_{2}$ with realistic defects.
We find that spin relaxation is highly sensitive to 
local symmetry and chemical bonds around defects. For the bilayer WSe$_2$, we identify
the scattering pathways in ultrafast dynamics and determine relevant
dynamical properties, essential to its utilization of unique spin-valley-layer
locking effects. Our work provides a predictive computational platform
for spin dynamics in solids, which has unprecedented potentials for
designing new materials ideal for spintronics and quantum information
technology.
\end{abstract}
\maketitle

\section{Introduction}

Spin is a fundamental quantum mechanical property of electrons and
other particles. The spin states can be used as the basis of quantum
bits in quantum information science (QIS)\citep{awschalom2018quantum},
in addition to being used in spintronics analogous to electrical charge
in conventional electronics\citep{vzutic2004spintronics}. The key
property for spintronics and spin-based QIS is the lifetime of spin
states. Stable manipulations of spin states in practical applications
require lifetimes on the order of hundreds of nanoseconds or even
milliseconds. Determining the underlying mechanisms and controlling
spin relaxation are vital to reach long spin lifetimes at room temperature.
Experimentally spin relaxation can be studied through ultrafast magneto-optical
pump-probe\citep{meier2012optical,dean2016ultrafast} and spin transport
measurements\citep{avsar2020colloquium}, allowing the direct observations
of dynamical processes and quantitative determination of spin relaxation
time, $\tau_{s}$.

Despite significant experimental progresses and several proposed systems
in the past decades\citep{pla2012single,vzutic2004spintronics}, materials
with properties required for practical QIS and spintronics applications
such as long $\tau_{s}$ at room temperature remain to be found\citep{awschalom2018quantum,avsar2020colloquium,RevModPhys.89.011001}.
Theoretical predictions of materials properties have been mostly focused
on electronic excitations\citep{reviewDreyer,Wu2017,Smart2018} and
electron-hole recombinations\citep{Wu2019,Wu20192,smart2020} of potential
spin defects for QIS applications. Reliable prediction of spin lifetime
and dominant relaxation mechanism will allow rational design of materials
in order to accelerate the identification of ideal materials for quantum
technologies, while forgoing the need of experimental search over
a large number of materials.

Until recently, most state-of-the-art theoretical methods to study
spin dynamics of solid-state materials are limited to simplified and
system-specific models that require prior input parameters \citep{vzutic2004spintronics,dyakonov1972spin,cummings2017giant,afanasiev2019ultrafast}.
These methods laid important theoretical foundation for spin dynamics,
such as the spin-Bloch kinetic equations developed from Non-equilibrium
Green's Function theory (NEGFT)\citep{wu2010spin}. Ref.\citenum{rosati2014derivation} derived a closed equation of motion for the electronic single-particle density matrix, including different scattering matrices, which may be applicable to spin dynamics. However, because
of the simplified electronic structure and electron-phonon coupling
matrices, quantitative prediction of spin relaxation remains out of
reach. Occasionally, even trends in $\tau_{s}$ predicted by such
models may be incorrect as shown recently for graphene.\citep{habib2020electric}
Furthermore, these models are unable to provide predictive values
for new materials where prior inputs are not available.

Prior to our work, the existing first-principles methodology for spin
lifetime has been mostly based on spin-flip matrix elements in a specialized
Fermi's Golden rule\citep{restrepo2012full,park2020spin,fedorov2013impact},
which is only applicable to systems with Kramers' degeneracy or spatial
inversion symmetry, not suitable to lots of materials promising for
quantum computing and spintronics applications~\citep{vzutic2004spintronics,avsar2020colloquium}.
Other first-principles techniques like real-time Time-Dependent Density
Functional Theory (TDDFT)\citep{marques2012fundamentals} are challenging
for crystalline systems due to high computational cost for describing
phonon relaxations that require large supercells. More importantly,
long simulation time over nanoseconds often required by spin relaxation
is a major difficulty for TDDFT, which is only practical for tens
to a few hundred femtoseconds. While spin dynamics based on TDDFT
has been recently performed for ultrafast demagnetization of magnetic
systems within tens of fs\citep{chen2019role,acharya2020ultrafast,krieger2015laser},
the intrinsic time scale and supercell limitations mentioned above
remain.

We recently derived a generalized rate equation~\citep{xu2020spin}
based on first-principles density matrix (DM) Lindblad dynamics framework,
which provides accurate spin relaxation time due to spin-orbit and
electron-phonon couplings for a broad range of materials, with arbitrary
symmetry. However, our previous work requires the system is already
at a quasi-equilibrium state when the dynamics can be described by
a single-exponential decay, which can not describe coupled spin and
carrier dynamics at an out-of-equilibrium state in ultrafast pump-probe
experiments. In this work, we develop a real-time \textit{ab initio} DM dynamics method
based on this theoretical framework, with complete 
descriptions of scattering processes including electron-phonon (e-ph),
electron-impurity (e-i), and electron-electron (e-e), being adequate
for over ns to ms simulation time. Specifically, compared to the generalized
rate equation in our previous work, DM dynamics with explicit real-time
evolutions allow coupled carrier and spin relaxation away from quasi-equilibrium
with all decoherence pathways simultaneously. This will facilitate
direct prediction of experimental signatures in ultrafast magneto-optical
spectroscopy to unambiguously interpret experimental probes of spin
and electron dynamics.

We will demonstrate the generality of our approach by considering two prototypical and disparate systems - GaAs and few-layer WSe$_2$, which have very different spin relaxation mechanisms.  

We will first apply our DM dynamics methodology
to investigate ultrafast spin dynamics of GaAs, which has broad interest
in spintronics over past decades\citep{vzutic2004spintronics,kikkawa1998resonant,hilton2002optical,jiang2009electron,kamra2011role}
and more recently\citep{PhysRevX.7.031010,PhysRevLett.121.033902,belykh2018quantum},
partly due to its long spin lifetime especially in the n-doped material
at relatively low temperature\citep{kikkawa1998resonant}. Despite
various experimental~\citep{kikkawa1998resonant,hilton2002optical,ohno1999spin,kimel2001room,hohage2006coherent}
and theoretical~\citep{vzutic2004spintronics,yu2005spin,jiang2009electron,mower2011dyakonov,kamra2011role,marchetti2014spin}
(mostly using parameterized model Hamiltonian) studies previously,
the dominant spin relaxation mechanism of bulk GaAs under various
temperatures and doping levels remains unclear. For example, Refs.\citenum{jiang2009electron}
and \citenum{mower2011dyakonov} claimed e-i and e-ph scatterings
dominate spin relaxation at low and room temperatures, respectively;
however, Refs.~\citenum{kamra2011role} and \citenum{marchetti2014spin}
conclude that e-e may be more important at room temperature and even
more at lower temperatures. Moreover, electron-phonon scattering matrices
which can be accurately obtained from first-principles, are very difficult
to be precisely described in parameterized models used previously.
Most importantly, the applicability of empirical D'yakonov-Perel'
(DP) relation, which is widely used for describing inversion-asymmetric
systems including GaAs, needs to be carefully examined. Throughout
this work, we provide complete and unambiguous insights on the underlying
mechanism of spin relaxation and applicability of the DP relation
for GaAs from first-principles DM dynamics.

Due to broken inversion symmetry and strong SOC,
monolayer transition metal dichalcogenides (TMDs) exhibit exciting
physical properties including  valley-specific optical excitation and spin-valley locking effects. In Ref. \citenum{dey2017gate,li2021valley},
it has been shown that by introducing doping in monolayer TMDs, ultraslow
decays of Kerr rotations, which correspond to ultralong spin/valley
lifetimes of resident carriers especially resident holes 
can be
observed at low temperatures. Those features make monolayer TMDs advantageous
for spin-valleytronics and (quantum) information processing.

Besides monolayers, bilayer TMDs recovering inversion symmetry
have also attracted significant interests because of the
new ``layer'' degree of freedom or layer pseudospin in addition
to spin and valley pseudospin\cite{gong2013magnetoelectric,xu2014spin,khani2020gate}.
Previous studies already concluded that electronic states at $K$/$K'$
valleys of a bilayer TMD are approximately a superposition of those of two monolayers. This allows us to tune which layer carriers/spins
are localized by a perpendicular electric field $E_{z}$, and make
use of the spin-valley-layer locking effects for spin-valleytronic
applications.

Although spin/valley relaxation of resident carriers
in monolayer TMDs, which is most relevant to  spin-valleytronic
applications, have been extensively examined\citep{xu2020spin,dey2017gate,li2021valley,song2016long}, the underlying 
dynamics especially the effects of different types of impurities
have not been investigated through predictive \emph{ab initio} simulations.
Furthermore, for bilayer TMDs, the study on spin/valley dynamics is still in infancy with few ultrafast measurements which however do not
exhibit long spin relaxation time and are lack of spin-valley-layer locking property \cite{guimaraes2018spin,ye2019ultrafast,bertoni2016generation}.
There is a lack of knowledge of the role of scattering processes and
the scattering pathways for spin/valley dynamics of free carriers
in bilayers, 
which prevents researchers to realize and manipulate spin-valley-layer locking effects for designing spin-valleytronic devices.

In this work, we will answer the above questions 
by performing \emph{ab initio} real-time dynamics simulations with
a circularly polarized pump pulse and relevant scattering mechanisms.
We focus on WSe$_{2}$ due to its larger valence band SOC splitting
and focus on dynamics of holes since $\tau_{s}$ of holes seem longer than electrons.

In the following, we first introduce our theoretical formalism of
real-time density-matrix approach with various scattering processes
and pump-probe spectroscopy. In particular, we focus on spin-orbit
mediated spin relaxation and decoherence processes under the existences
of electron scatterings, which are rather common in semiconductors
and metals\citep{wu2010spin,vzutic2004spintronics}. We use this method
to simulate pump-probe Kerr rotation and real-time spin dynamics,
by using GaAs as a prototypical example and comparing with experiments.
Next, we study spin lifetime dependence on the temperature and doping
level, where the dominant mechanisms can vary significantly. We then
discuss the roles of different scattering mechanisms and phonon modes
in carrier and spin relaxations, respectively, in order to resolve
related long-standing controversies. We further simulate
ultrafast dynamics in monolayer and bilayer WSe$_{2}$ and extract
useful dynamical properties. Our work provides predictive theory
and computational platform for open quantum dynamics, and offers new
and critical insights for spin relaxation and decoherence in general
solid-state systems.

\section{Theory}

\subsection{Real-time density-matrix dynamics and spin relaxation time\label{subsec:theory_dynamics}}

To provide a general formulation of quantum dynamics in solid-state
materials, we start from the Liouville-von Neumann equation in the
interaction picture, 
\begin{align}
\frac{d\rho\left(t\right)}{dt} & =-i[H'\left(t\right),\rho\left(t\right)],\label{eq:Liouville}\\
H'\left(t\right) & =H\left(t\right)-H_{0}\left(t\right),
\end{align}
where $H$, $H_{0}$ and $H'$ are total, unperturbed and perturbed
Hamiltonian, respectively. In this work, the total Hamiltonian is
\begin{align}
H= & H_{0}+H\sub{pump}+H\sub{e-i}+H\sub{e-ph}+H\sub{e-e},\label{eq:H}\\
H_{0}= & H_{e,0}+H\sub{efield}+H_{z}+H\sub{ph},
\end{align}
where $H_{e,0}$ is electronic Hamiltonian under zero external field.
In this work, $H\sub{efield}$ is Hamiltonian induced by a perpendicular electric field $E_z$ along the vacuum direction.
$H_{z}$ is the Zeeman Hamiltonian corresponding
to an external magnetic field $\vec{B}$, $H_{z}=g_{s}\mu_{B}\vec{B}\cdot\vec{s}$,
where $\vec{s}=\left(s_{x},s_{y},s_{z}\right)$ and $s_{i}$ is spin
matrix in Bloch basis under zero field. $g_{s}$ is $g$ factor and
$\mu_{B}$ is the Bohr magneton. $H\sub{pump}$ is the Hamiltonian
of the pump pulse and will be described below. $H\sub{ph}$ is the
phonon Hamiltonian, while $H\sub{e-i}$, $H\sub{e-ph}$ and $H\sub{e-e}$
describe the electron-impurity, electron-phonon and electron-electron
interactions respectively. The detailed forms of the interaction Hamiltonians
are given in Appendix A.

In practice, the many-body density matrix master equation in Eq.~\ref{eq:Liouville}
is reduced to a single-particle one and the environmental degrees
of freedom are traced out\citep{rossi2002theory}. The total rate
of change of the density matrix is separated into terms related to
different parts of Hamiltonian, 
\begin{align}
\frac{d\rho}{dt}= & \frac{d\rho}{dt}|\sub{coh}+\frac{d\rho}{dt}|\sub{scatt},\label{eq:dynamics}
\end{align}
where $\rho$ is the density matrix of electrons. Above, $\frac{d\rho}{dt}|\sub{coh}$
describes the coherent dynamics of electrons under potentials or fields,
e.g. the applied pump pulse, while $\frac{d\rho}{dt}|\sub{scatt}$
captures the scattering between electrons and other particles.

To obtain Eq. \ref{eq:dynamics} which involves only
the dynamics of electrons or the electronic subsystem, we have assumed
the environmental subsystem is not perturbed by the change of the electronic subsystem, 
which in this work means there is no dynamics of phonons. 
This assumption is valid when the system is not far from equilibrium, e.g.,
when excitation is weak.
In most spin dynamics experiments, it is desirable to work in the low excitation density limit to avoid additional complexities and focus on the physics of spin dynamics.
Indeed in many experiments, e.g., in Refs. \citenum{kikkawa1998resonant} and \citenum{yang2015long}, pump fluence and excitation density are controlled to be low, e.g., excitation density 2$\times$10$^{14}$ cm$^{-2}$ for GaAs.
Therefore, phonon dynamics can be safely excluded in the current stage.
The inclusion of phonon degrees of freedom in the density-matrix dynamics has been discussed 
in detail in Refs. \citenum{iotti2017phonon,rossi2002theory} with model Hamiltonian, which can be our future work to implement from first-principles.

To define spin lifetime, we follow the time evolution of the observable
\begin{align}
S_{i} & =\mathrm{Tr}\left(s_{i}\rho\right),\label{eq:spin_observable}
\end{align}
where $s_{i}$ is the spin operator ($i=x,y,z$). This time evolution
must start at an initial state (at $t=t_{0}$) with a net spin i.e.
$\delta\rho(t_{0})=\rho(t_{0})-\rho\super{eq}\neq0$ such that $\delta S_{i}(t_{0})=S_{i}\left(t_{0}\right)-S_{i}\super{eq}\neq0$,
where ``eq'' corresponds to the final equilibrium state. We evolve
the density matrix through Eq.~\ref{eq:dynamics} using an adaptive
Runge-Kutta fourth-order method for a long enough simulation time,
typically from tens of ps to several ns, until the evolution of $S_{i}\left(t\right)$
can be reliably fitted by 
\begin{align}
S_{i}\left(t\right)-S_{i}\super{eq}= & \left[S_{i}\left(t_{0}\right)-S_{i}\super{eq}\right]exp\left[-\frac{t-t_{0}}{\tau_{s,i}}\right]\nonumber \\
 & \times\mathrm{cos}\left[\omega_{B}\left(t-t_{0}\right)+\phi\right].
\end{align}
to extract the relaxation time, $\tau_{s,i}$. Above, $\omega_{B}$
is oscillation frequency due to energy splitting in general, which
under an applied magnetic field $\vec{B}$ would include a contribution
$\approx0.5g_{s}\mu_{B}\left(\vec{B}\times\hat{\vec{S_{i}}}\right)$.

In order to examine whether the spin relaxation time depends on how
the spin imbalance is generated, we implement two general ways to
initialize $\delta\rho(t_{0})$. First, for simulating pump-probe
experiments, we choose $\delta\rho(t_{0})$ corresponding to interaction
with a pump pulse. Second, we use the technique proposed previously
in Ref. \citenum{xu2020spin} by applying a test magnetic field
at $t=-\infty$, allowing the system to equilibrate with a net spin
and then turning it off suddenly at $t_{0}$.

\subsection{Scattering terms}

The scattering part of the master equation can be separated into contributions
from several scattering channels, 
\begin{align}
\frac{d\rho}{dt}|\sub{scatt}= & \sum_{c}\frac{d\rho}{dt}|_{c},
\end{align}
where $c$ labels a scattering channel. Under Born-Markov approximation,
in general we have\citep{rosati2014derivation} 
\begin{align}
\frac{d\rho_{12}}{dt}|_{c}= & \frac{1}{2}\sum_{345}\left[\begin{array}{c}
\left(I-\rho\right)_{13}P_{32,45}^{c}\rho_{45}\\
-\left(I-\rho\right)_{45}P_{45,13}^{c,*}\rho_{32}
\end{array}\right]+H.C.,\label{eq:scattering}
\end{align}
where $P^{c}$ is the generalized scattering-rate matrix and H.C.
is Hermitian conjugate. The subindex, e.g., ``1'', is the combined
index of k-point and band. The weights of k points must be considered
when doing sum over k points. Note that $P^{c}$ in the interaction
picture is related to its value $P^{S,c}$ in the Schrodinger picture
as 
\begin{align}
P_{1234}^{c}\left(t\right)= & P_{1234}^{S,c}\mathrm{exp}\left[i\left(\epsilon_{1}-\epsilon_{2}-\epsilon_{3}+\epsilon_{4}\right)t\right],
\end{align}
where $\epsilon_{i}$ are single-particle eigenvalues of $H_{0}$.
Below, we consider three separate scattering mechanisms - electron-impurity
(e-i), electron-phonon (e-ph) and electron-electron (e-e), and describe
the matrix elements for each.

For electron-phonon scattering, the scattering matrix is given by\citep{rosati2014derivation}
\begin{align}
P_{1234}\super{S,e\text{-}ph}= & \sum_{q\lambda\pm}A_{13}^{q\lambda\pm}A_{24}^{q\lambda\pm,*},\\
A_{13}^{q\lambda\pm}= & \sqrt{\frac{2\pi}{\hbar}}g_{12}^{q\lambda\pm}\sqrt{\delta_{\sigma}^{G}\left(\epsilon_{1}-\epsilon_{2}\pm\omega_{q\lambda}\right)}\sqrt{n_{q\lambda}^{\pm}},
\end{align}
where $q$ and $\lambda$ are phonon wavevector and mode, $g^{q\lambda\pm}$
is the electron-phonon matrix element, resulting from the absorption
($-$) or emission ($+$) of a phonon, computed with self-consistent
spin-orbit coupling from first-principles,\citep{giustino2017electron}
$n_{q\lambda}^{\pm}=n_{q\lambda}+0.5\pm0.5$ in terms of phonon Bose
factors $n_{q\lambda}$, and $\delta_{\sigma}^{G}$ represents an
energy conserving $\delta$-function broadened to a Gaussian of width
$\sigma$.

Next, for electron-impurity scattering, the scattering matrix is given
by 
\begin{align}
P_{1234}\super{S,e\text{-}i}= & A_{13}^{i}A_{24}^{i,*},\\
A_{13}^{i}= & \sqrt{\frac{2\pi}{\hbar}}g_{13}^{i}\sqrt{\delta_{\sigma}^{G}\left(\epsilon_{1}-\epsilon_{3}\right)}\sqrt{n_{i}V\sub{cell}},\\
g_{13}^{i}= & \left\langle 1\right|V^{i}\left|3\right\rangle ,
\end{align}
where $n_{i}$ and $V\sub{cell}$ are impurity density and unit cell
volume, respectively, and $V^{i}$ is the impurity potential. In this
work, we deal with ionized and neutral  impurities differently. For ionized impurities, $V^{i}$ is proportional
to screened Coulomb potential\citep{jacoboni2010theory}; for neutral impurities, we compute impurity potentials with supercell methods from DFT. (See Appendix
A for further details).

Finally, for electron-electron scattering, the scattering matrix is
given by\citep{rosati2014derivation} 
\begin{align}
P_{12,34}\super{S,e\text{-}e}= & 2\sum_{56,78}\left(I-\rho\right)_{65}\mathscr{A}_{15,37}\mathscr{A}_{26,48}^{*}\rho_{78},\\
\mathscr{A}_{1234}= & \frac{1}{2}\left(A_{1234}-A_{1243}\right),\\
A_{1234}= & \frac{1}{2}\sqrt{\frac{2\pi}{\hbar}}\left[g_{1234}^{e-e}(\delta_{\sigma,1234}^{G})^{1/2}+g_{2143}^{e-e}(\delta_{\sigma,2143}^{G})^{1/2}\right],\\
g_{1234}\super{e\text{-}e}= & \left\langle 1\left(r\right)\right|\left\langle 2\left(r'\right)\right|V\left(r-r'\right)\left|3\left(r\right)\right\rangle \left|4\left(r'\right)\right\rangle ,
\end{align}
where $V\left(r-r'\right)$ is the screened Coulomb potential and
$\delta_{\sigma,1234}^{G}=\delta_{\sigma}^{G}\left(\epsilon_{1}+\epsilon_{2}-\epsilon_{3}-\epsilon_{4}\right)$
is a Gaussian-broadened energy conservation function. The screening
is described by Random-Phase-Approximation (RPA) dielectric function
(details in Appendix A). Although the above equations describe all
possible scattering processes between electrons and holes, we only
consider those between conduction electrons here, which are appropriate
for n-type Group III-V semiconductors\citep{jiang2009electron,mower2011dyakonov}.
The electron-hole scattering can be important for intrinsic and p-type
material.\citep{jiang2009electron,mower2011dyakonov} We note that
unlike the e-ph and e-i channels, $P\super{S,e\text{-}e}$ (as well as the dielectric screening in $V$) is a function
of $\rho$ and needs to be updated during time evolution of $\rho$.
This is a clear consequence of the two-particle nature of e-e scattering.
$P\super{S,e\text{-}e}$ can be written as the difference between
a direct term and an exchange term, 
\begin{align}
P\super{S,e\text{-}e}= & P\super{S,e\text{-}e,d}-P\super{S,e\text{-}e,x},\\
P\super{S,e\text{-}e,d}= & \sum_{56,78}\left(I-\rho\right)_{65}A_{15,37}A_{26,48}^{*}\rho_{78},\\
P\super{S,e\text{-}e,x}= & \sum_{56,78}\left(I-\rho\right)_{65}A_{15,37}A_{26,84}^{*}\rho_{78}.
\end{align}
According to Ref. \citenum{rossi2002theory}, the direct term is
expected to dominate the dynamical scattering processes between conduction
or valence electrons, allowing us to neglect the exchange term here.

\subsection{Pump-probe simulation}

In nonrelativistic limit, the light-matter interaction
Hamiltonian operator ($\widehat{H}_{e-p}$) reads\cite{joly2009interaction}
\begin{align*}
\widehat{H}_{e-p}= & \frac{e}{m_{e}}\vec{A}\left(t\right)\cdot\widehat{\vec{p}}\\
 & +\frac{e}{2m_{e}}\vec{A}\left(t\right)\cdot\vec{A}\left(t\right)+g_{e}\mu_{B}\widehat{\vec{s}}\cdot\left(\bigtriangledown\times\vec{A}\left(t\right)\right),
\end{align*}
where $\vec{A}\left(t\right)$ is the vector potential
and $\vec{A}\left(t\right)=\vec{A}_{0}\left(t\right)e^{-i\omega t}+\vec{A}_{0}^{*}\left(t\right)e^{i\omega t}$
with $\vec{A}_{0}\left(t\right)$ being the complex amplitude and
$\omega$ being photon frequency. $\widehat{\vec{p}}$ is momentum
operator. $g_{e}\approx2.0023192$ is anomalous gyromagnetic ratio.
The second quadratic term plays a role 
only when pump fluence is higher by several orders of magnitude 
than that in usual spin dynamics experiments and can be safely neglected.
Since $\bigtriangledown\times\vec{A}\left(t\right)=-i\vec{q}_{\mathrm{photon}}\times\vec{A}\left(t\right)$\cite{joly2009interaction}
and the photon wavevector $\vec{q}_{\mathrm{photon}}$ is quite small (the photon wavelength is much longer than the scale of unit cells),
the third term is also negligible. Therefore, we will only keep the
first $\vec{A}\left(t\right)\cdot\widehat{\vec{p}}$ term.

The interaction with a pump pulse of frequency $\omega\sub{pump}$
in the interaction picture is given by 
\begin{align}
H_{\mathrm{pump},k,mn}\left(\omega\sub{pump},t\right)= & \frac{e}{m_{e}}\vec{A}_{0}\left(t\right)\cdot\vec{p}_{k,mn}e^{it\left(\epsilon_{m}-\epsilon_{n}-\omega\sub{pump}\right)}\nonumber \\
 & +H.C.,
\end{align}
where $m,n$ represent the band indices and $k$ represents the k point
sampling in the first Brillouin zone. For a Gaussian pulse centered
at time $t\sub{center}$ with width $\tau\sub{pump}$, 
\begin{align}
\vec{A}_{0}\left(t\right) & =\vec{A}_{0}\frac{\mathrm{1}}{\sqrt{\sqrt{\pi}\tau\sub{pump}}}\mathrm{exp}\left[-\left(t-t\sub{center}\right)^{2}/\left(2\tau\sub{pump}^{2}\right)\right].
\end{align}
Note that the corresponding pump fluence is $I\sub{pump}=\omega\sub{pump}^{2}|A_{0}|^{2}/\left(8\pi\alpha\right)$,
where $\alpha$ is fine structure constant. As a part of the coherent
portion of the time evolution, the dynamics due to this term are captured
directly in the Liouville form,\citep{d2020real,hannewald2000quantum}
\begin{align}
\frac{d\rho}{dt}|\sub{pump}= & -i[H\sub{pump},\rho].
\end{align}

The probe pulse interacts with the material similarly to the pump
pulse, and could be described in exactly the same way in principle.
However, this would require repeating the simulation for several values
of the pump-probe delay. Instead, since the probe is typically chosen
to be of sufficiently low intensity, we use second-order time-dependent
perturbation theory to capture its interaction with the system,

\begin{align}
\Delta\rho\super{probe}= & \frac{1}{2}\sum_{345}\left\{ \begin{array}{c}
\left[I-\rho\left(t\right)\right]_{13}P_{32,45}\super{probe}\rho\left(t\right)_{45}\\
-\left[I-\rho\left(t\right)\right]_{45}P_{45,13}\super{probe,*}\rho\left(t\right)_{32}
\end{array}\right\} +H.C.,\label{eq:drho_probe}
\end{align}
where $P\super{probe}$ is the generalized scattering-rate matrix
for the probe in the interaction picture. Its corresponding Schrodinger-picture
quantity is 
\begin{align}
P_{1234}\super{S,probe}= & \sum_{\pm}A_{13}\super{probe,\pm}A_{24}\super{probe,\pm,*},\\
A_{13}\super{probe,\pm}= & \sqrt{\frac{2\pi}{\hbar}}\frac{e}{m_{e}}\left(\vec{A}_{0}\super{probe}\cdot\vec{p}\right)\sqrt{\delta_{\sigma}^{G}\left(\epsilon_{1}-\epsilon_{3}\pm\omega\sub{probe}\right)}.
\end{align}
The dielectric function change $\Delta\epsilon$ between the excited
state and ground state absorption detected by the probe is then 
\begin{align}
\mathrm{Im}\Delta\epsilon= & \frac{2\pi}{\left(\omega\sub{probe}\right)^{3}|A_{0}\super{probe}|^{2}}\mathrm{Tr}\left(H_{0}\Delta\rho\super{probe}\right).
\end{align}
Note that $\Delta\rho\super{probe}$ contains $|A_{0}\super{probe}|^{2}$
so that $\mathrm{Im}\Delta\epsilon$ is independent of $A_{0}\super{probe}$.
The above $\mathrm{Im}\Delta\epsilon$ is a functional of the density
matrix according to Eq.~\ref{eq:drho_probe} and is an extension
of the usual independent-particle $\mathrm{Im}\epsilon$ depending
on just occupation numbers.\citep{molina2017ab} After computing $\mathrm{Im}\Delta\epsilon$
above, the real part $\mathrm{Re}\Delta\epsilon$ can be obtained
from the Krames-Kronig relation.

By summing up the dielectric function change $\Delta\epsilon$ computed
above with the dielectric function for ground state absorption, we
can obtain the excited-state $\epsilon$ as inputs for Kerr and Faraday
rotation calculations.\citep{mainkar1996first} These correspond to
the rotations of the polarization plane of a linearly polarized light,
reflected by (Kerr) and transmitted through (Faraday) the material,
after a pump excitation with a circularly-polarized light. Time-Resolved
Kerr/Faraday Rotation (TRKR/TRFR) has been widely used to study spin
dynamics of materials\citep{kikkawa1998resonant,kimel2001room}. In
a TRKR experiment, a circularly-polarized pump pulse is used to excite
valence electrons of the sample to conduction bands. The transitions
approximately satisfy the selection rule of $\Delta m_{j}=\pm1$ for
left and right circularly-polarized pulses, respectively, where $m_{j}$
is secondary total angular momentum. TRKR works by measuring the changes
of polarization of reflected light, which qualitatively is proportional
to the small population imbalance of electronic states with different
$m_{j}$.

Specifically, the Kerr rotation angle $\theta_{K}$ is computed with
dielectric functions by 
\begin{align}
\theta_{K}= & \mathrm{Im\frac{\sqrt{\epsilon_{+}}-\sqrt{\epsilon_{-}}}{1-\sqrt{\epsilon_{+}}\sqrt{\epsilon_{-}}}},\label{eq:kerr}
\end{align}
where $\pm$ denotes the left and right circular polarization, respectively.

\section{Computational details}

The ground-state electronic structure, phonon, and e-ph matrix element
calculations of GaAs and few-layer WSe$_{2}$ are first calculated
using Density Functional Theory (DFT) with relatively coarse $k$
and $q$ meshes in the JDFTx plane-wave DFT code.\citep{sundararaman2017jdftx}
For GaAs, we use the experimental lattice constant of 5.653~$\text{Å}$,\citep{GaAs_lattice}
and select the SCAN exchange-correlation functional\citep{sun2015strongly}
for an accurate description of the electron effective mass (see section
II in Supplemental Materials\citep{supplementalmaterial}). We also
apply a scissor operator to the DFT values to reach experimental band
gap 1.43 eV\citep{madelung1987semiconductors}. 
For WSe$_{2}$, we used PBE exchange correlation functional along with
the DFT-D2 pair potential dispersion corrections\citep{grimme2006semiempirical}.
The resulting 
lattice constant is 3.32 $\text{Å}$ and distance between two W-atom planes is 6.419~$\text{Å}$
close to experimental values of bulk WSe$_2$, 3.297 and 6.491~$\text{Å}$\citep{agarwal1979growth}.
The phonon calculations
of GaAs and WSe$_{2}$ employ a $4\times4\times4$ and $6\times6$
supercell, respectively. We use Optimized Norm-Conserving Vanderbilt
(ONCV) pseudopotentials\citep{hamann2013optimized} with self-consistent
spin-orbit coupling throughout, which we find converged at a plane-wave
kinetic energy cutoff of 34 and 62 Ry for GaAs and WSe$_{2}$, respectively.
With these computational parameters, we find the effective mass of
conduction electrons of GaAs to be 0.054$m_{e}$, close to the experimental
value of 0.067$m_{e}$\citep{madelung1987semiconductors}. (More convergence
tests can be found in Supporting  Information (SI)\citep{supplementalmaterial}).

We then transform all quantities from plane wave basis to maximally
localized Wannier function basis\citep{marzari1997maximally}, and
interpolate them\citep{PhononAssisted,giustino2017electron,GraphiteHotCarriers,brown2017experimental,NitrideCarriers,TAparameters}
to substantially finer k and q meshes. The Wannier interpolation approach
fully accounts for polar terms in the e-ph matrix elements and phonon
dispersion relations, using the approach developed by Verdi and Giustino\citep{verdi2015frohlich} for 3D and using the methods in Ref.~\citenum{Sohier2016} and Ref.~\citenum{Sohier2017} for 2D systems.
The Born effective charges and dielectric constants are calculated
from open-source code QuantumESPRESSO\citep{giannozzi2009quantum}.

For GaAs, the fine $k$ and $q$ meshes are $288\times288\times288$
for simulations at 300 K and are finer at lower temperature, e.g.,
$792\times792\times792$ for simulations at 30 K. This is necessary
to sample enough electronic states around band edges and for spin
lifetime convergence within 20$\%$. The $k$ and $q$ convergence
are easier for WSe$_{2}$ due to much larger effective masses and
we used $168\times168$ and $600\times600$ meshes at 50 and 10 K,
respectively. The computation of e-i and e-e matrix elements and the
real-time dynamics simulations are done with a new custom code interfaced
to JDFTx. The energy-conservation smearing parameter $\sigma$ is
chosen to be comparable or smaller than $k_{B}T$ for each calculation.
Detailed convergence tests of number of k points and energy window
for electronic states at various smearing parameters can be found
in Supplemental Materials\citep{supplementalmaterial}.

\section{Results and discussions\label{sec:results}}

\subsection{Applications to n-doped GaAs}

\subsubsection{Spin dynamics and its relation to TRKR\label{subsec:GaAs_ultrafast}}

\begin{figure}
\includegraphics[scale=0.4]{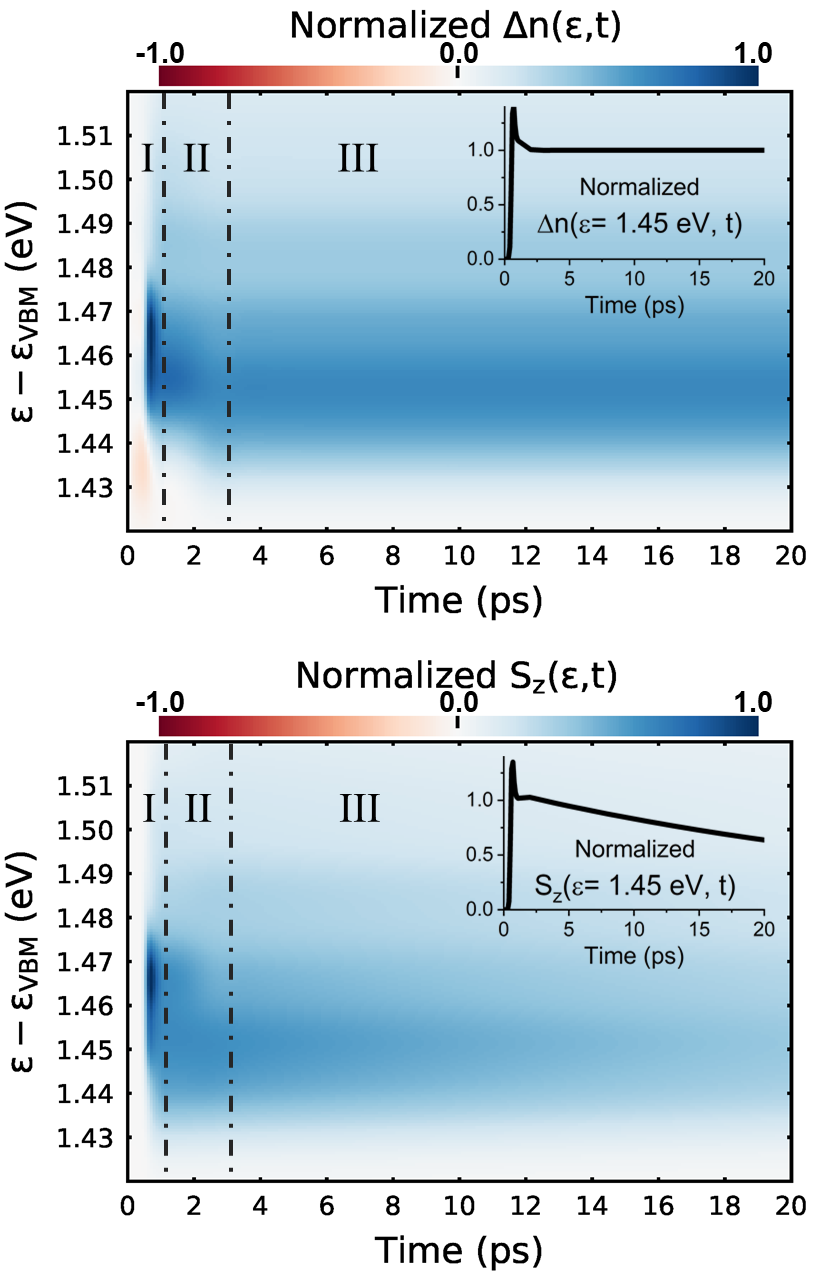} \caption{The energy-resolved dynamics of carriers (a) $\Delta n\left(\epsilon,t\right)\,=\,n\left(\epsilon,t\right)-n\left(\epsilon,0\right)$
and (b) spins $S_{z}\left(\epsilon,t\right)$ of conduction electrons
with a circularly polarized pump pulse centered
at 0.5 ps. The insets on the top right of both panels show $\Delta n\left(\epsilon,t\right)$
and $S_{z}\left(\epsilon,t\right)$ at $\epsilon$=1.45 eV. The pump
energy $\omega\sub{pump}$=1.47 eV is chosen to be higher than band
gap 1.43 eV\citep{madelung1987semiconductors}. The width of the pump
pulse $\tau\sub{pump}$ is 100 fs. The pump fluence $I\sub{pump}$ is
low at 0.01 $\mathrm{\mu J}$ cm$^{-2}$. The dynamics can be approximately
divided into three regions - Region I, II and III labeled in this
figure. In Region I, the system is excited by a pump pulse. In Region
II, pump processes are already finished, then both carriers and spins
relax simultaneously. In Region III, carrier distribution stays unchanged
while spins keep decaying.\label{fig:energy-resolved-dynamics}}
\end{figure}

In general, time evolution of Kerr rotation angle $\theta_{K}$ (see
Eq. \ref{eq:kerr}) is not equivalent to that of spin along the direction
of reflected light, and in fact, they can be quite different in some
cases\citep{dey2017gate}. There are few first-principles studies
of TRKR considering scattering processes in a form of semiclassical
Boltzmann equation\citep{molina2017ab}. A full quantum description
of scatterings with non-diagonal density matrix in TRKR has not been
presented in previous first-principles studies, to the best of our
knowledge. And the relation between dynamics of $\theta_{K}$ and
spin observable for general systems including GaAs has not yet been
well examined.

Using our density-matrix approach, we are able to directly simulate
the nonequilibrium ultrafast dynamics of optically excited systems
during which the dynamics of different electronic quantities such
as spin and carriers can be strongly coupled. We include all scattering
terms in a full quantum description as shown in the theory section
II.B and Appendix A. We perform the real-time dynamics simulations
of n-type GaAs for tens of ps at room temperature and several ns at
low temperature until the fitted spin lifetime does not change any
more. Having temporal density matrix, we can further analyze the dynamics
of various observables, including occupation, spin and Kerr rotation
angle easily. We then examine the relation between $\theta_{K}$ and
spin in the dynamics.

\begin{figure}
\includegraphics[scale=0.25]{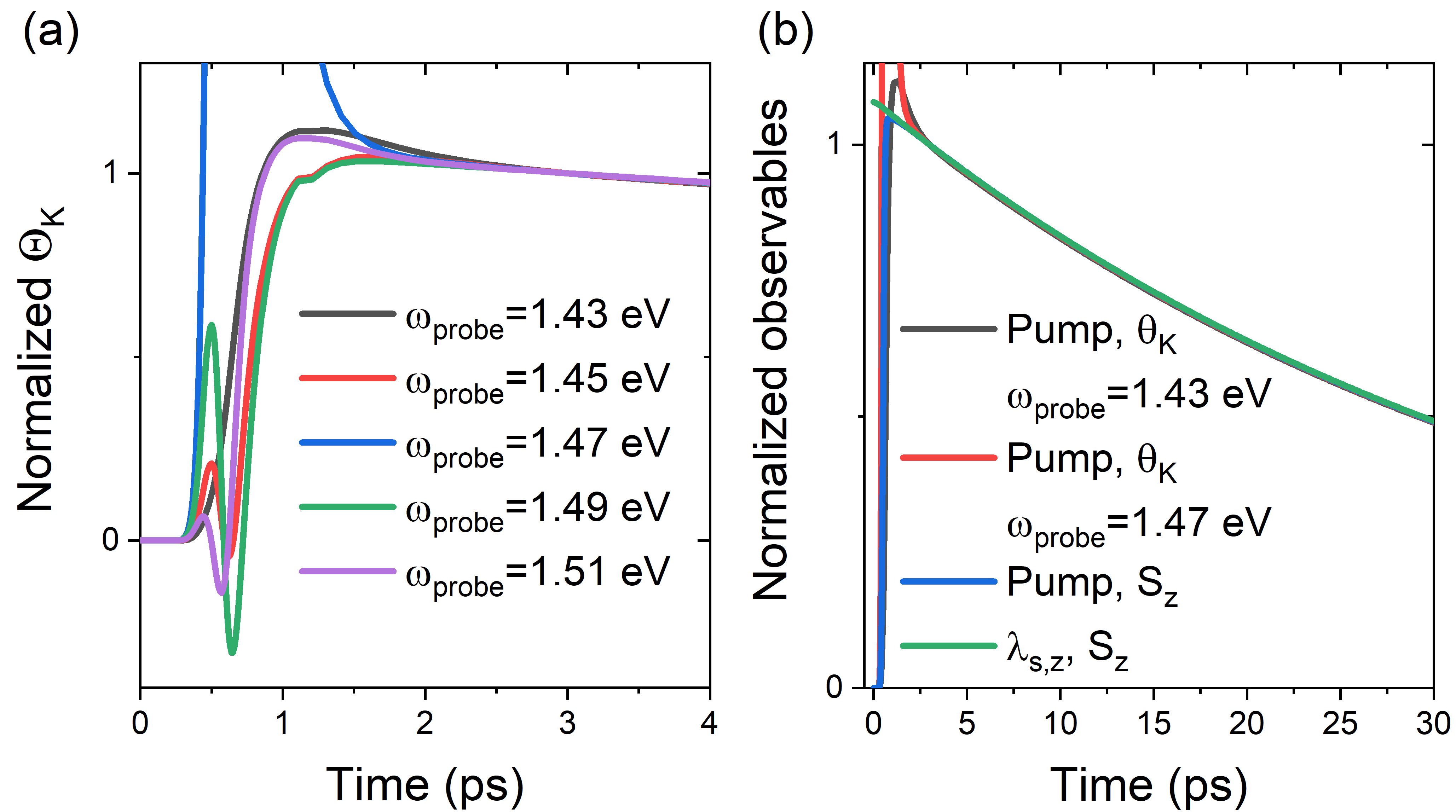} \caption{(a) Compare the dynamics of Kerr rotation angle $\theta_{K}$ at different
probe energies $\omega\sub{probe}$ excited by a circularly
polarized pump pulse in first few ps. Fast oscillations in first
2 ps are due to the pump pulse and coupled spin and carrier relaxation.
(b) Compare relaxation of different observables - $\theta_{K}$ with
different $\omega\sub{probe}$ (denoted by black and red lines) and
$S_{z}$ with initial spin imbalance generated by a pump pule (Pump)
or a test magnetic field along z direction $\lambda_{s,z}\sim0.001-0.1$
Tesla (blue and green lines). $\omega\sub{pump}$=1.47 eV. The pump
pulse is centered at 0.5 ps. The longer time-scale dynamics (over
10 ps) for $\theta_{K}$ and $S_{z}$ has similar relaxation time
independent on the generation method of spin imbalance and specific
probe energies. 
\label{fig:spin_generators_and_observables}}
\end{figure}

Figure \ref{fig:energy-resolved-dynamics} shows the energy-resolved
dynamics of carriers $\Delta n\left(\epsilon,t\right)\,=\,n\left(\epsilon,t\right)-n\left(\epsilon,0\right)$
and spins $S_{z}\left(\epsilon,t\right)$. The energy-resolved observable
$O\left(\epsilon\right)$ is defined as $\mathrm{Re}\left[\sum_{k,mn}o_{k,mn}\rho_{k,nm}\delta\left(\epsilon-\epsilon_{km}\right)\right]$,
where $o$ is operator matrix. We can see that during the first ps
(region I in Fig.~\ref{fig:energy-resolved-dynamics}), both observables
vary quickly due to the existence of the pump processes and both have
their maximum at an energy slightly lower than the pump energy, 1.47
eV (slightly larger than the band gap 1.43 eV), at a time shortly
after the time center of the pump pulse - 0.5 ps. Interestingly, after
pump being not active or after 0.8-1 ps, carriers and spins simultaneously
relax until 2-3 ps (region II in Fig.~\ref{fig:energy-resolved-dynamics}a
and \ref{fig:energy-resolved-dynamics}b). Afterward (region III in
Fig.~\ref{fig:energy-resolved-dynamics}a and \ref{fig:energy-resolved-dynamics}b),
carriers stay unchanged but spins $S_{z}\left(\epsilon,t\right)$
decay exponentially as shown in the insets of Fig. \ref{fig:energy-resolved-dynamics}a
and \ref{fig:energy-resolved-dynamics}b.

We have further analyzed the dynamics of Kerr rotation angle $\theta_{K}$
and compared it with spin dynamics. From Fig.~\ref{fig:spin_generators_and_observables}a,
we can see that during pump processes and shortly after them (from
0 to 2 ps), $\theta_{K}\left(t\right)$ has strong oscillations and
sensitive to the probe energy $\omega\sub{probe}$. The $\omega\sub{probe}$-sensitivity
may be partly attributed to the energy dependence of carrier and spin
dynamics. From Fig.~\ref{fig:spin_generators_and_observables}a and
\ref{fig:spin_generators_and_observables}b, it can be seen that after
3 ps (or in time region III defined in Fig.~\ref{fig:energy-resolved-dynamics}),
$\theta_{K}$ with different $\omega\sub{probe}$ decay exactly the
same. We can also find that with a pump pulse, relaxation time of
the Kerr rotation is the same as that of $S_{z}$, i.e. $\tau_{s,z}$.
Moreover, it turns out that $\tau_{s,z}$ does not depend on how spin
imbalance is generated - by a circularly polarized
pump pulse or by turning off a test magnetic field along $z$ direction
(see Sec. \ref{subsec:theory_dynamics}). This may indicate that if
the system is not extremely far from equilibrium, spin relaxation
along direction $i$ is not sensitive to the way of generating spin
imbalance, as long as the degrees of freedom other than $S_{i}$ are
not relevant or disappear in a short time. According to these observations,
hereinafter, we will do real-time dynamics starting from a $\delta\rho$
generated by turning off a test magnetic field and fit $\tau_{s,z}$
from time evolution of $S_{z}$.

\begin{figure}
\includegraphics[scale=0.35]{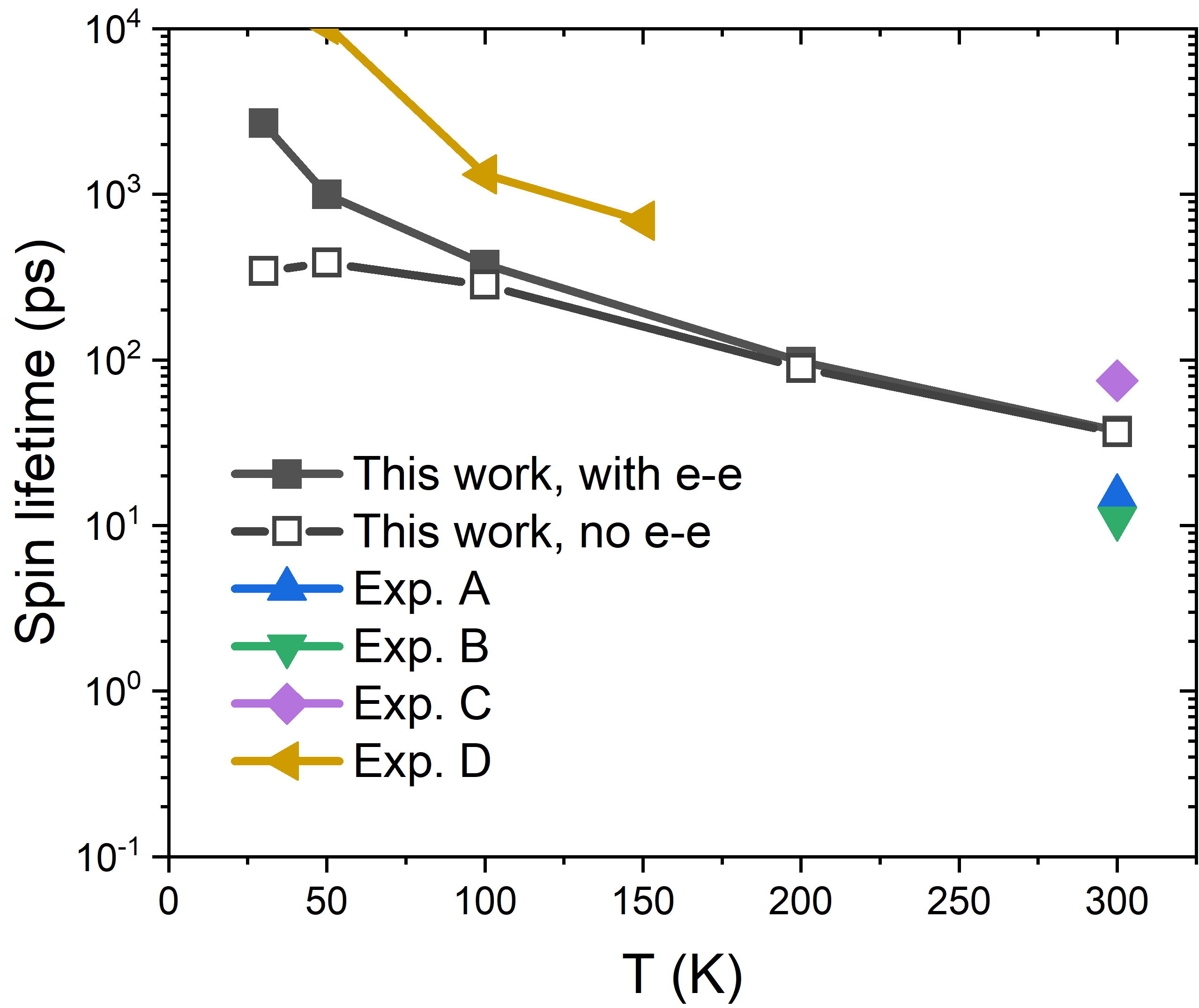} \caption{Theoretical spin lifetime with (black solid square) and without (black
empty square) the electron-electron scattering compared with experimental
data. Exp. A, B, C and D are experimental data from Refs. \citenum{kimel2001room,bungay1997direct,hohage2006coherent}
and \citenum{kikkawa1998resonant}, respectively.\label{fig:compare_exp}}
\end{figure}

We have also studied the effects of $\omega\sub{pump}$ and pump fluence
$I\sub{pump}$ on spin relaxation of $n$-GaAs at 300 K. We find that
$\omega\sub{pump}$ has very weak effects on spin relaxation but $\tau_{s,z}$
decreases with pump fluence. See more details in Appendix C.

\subsubsection{Temperature-dependence of spin lifetime and its dominant relaxation
mechanism\label{subsec:temperature}}

As discussed earlier, long-standing controversies remain for the dominant
spin relaxation mechanism of GaAs at different temperature and doping
level~\citep{jiang2009electron, mower2011dyakonov, kamra2011role, marchetti2014spin},
which will be resolved in the following sections. We start from study
$\tau_{s,z}$ of $n$-GaAs as a function of temperature at a moderate
doping level ($2\times10^{16}$ cm$^{-3}$). For simplicity, we assume
all impurities are fully ionized, so that the impurity density $n_{i}$
is equal to the free carrier density $n\sub{free}$. We first compared
our calculated spin lifetime with experimental results in Fig. \ref{fig:compare_exp}.
Our results of $\tau_{s,z}$ of $n$-GaAs give good agreement with
experiments at various temperatures\citep{kimel2001room,bungay1997direct,hohage2006coherent,kikkawa1998resonant}.
Different experiments have slight variations between each other due
to sample preparation conditions and specific measurement techniques.
The spin lifetime increases from tens of ps at room temperature to
tens of ns at low temperature. Note that e-e scattering plays an essential
role at low temperatures, i.e. by comparing with (black solid square)
and without (black empty square) in Fig.~\ref{fig:compare_exp}.
The correct temperature dependence of $\tau_{s,z}$ can be reproduced
only if e-e scattering is included.

\begin{figure}
\includegraphics[scale=0.35]{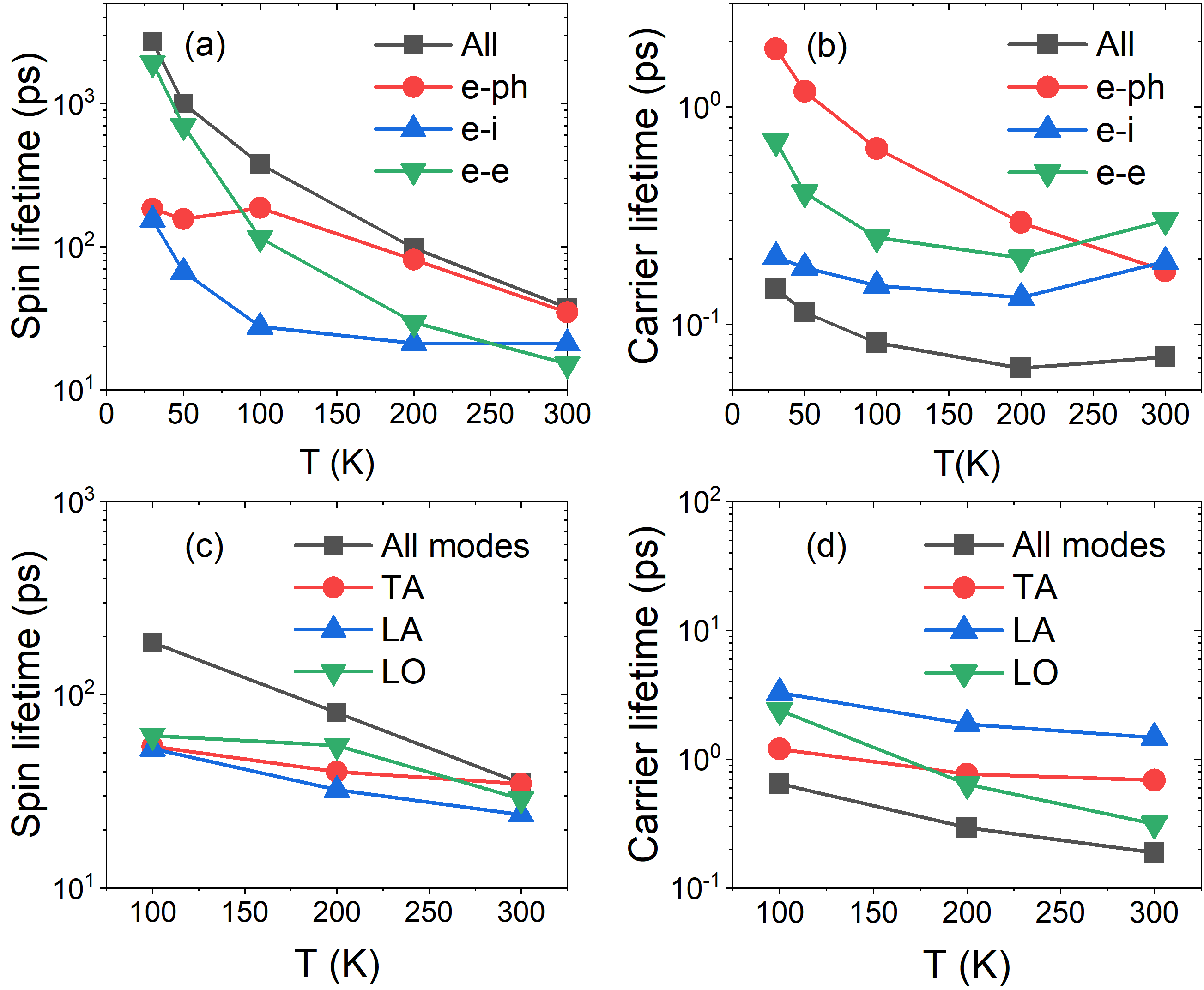} \caption{Spin and carrier lifetimes of $n$-GaAs with $n_{i}=2\times10^{16}$
cm$^{-3}$ with different scattering mechanisms and different phonon
modes. In (a) and (b), ``All'' represents all the e-ph, e-i and
e-e scattering mechanisms being considered. TA, LA and LO represent
transverse acoustic, longitudinal acoustic and longitudinal optical
modes, respectively. The carrier lifetimes $\overline{\tau}_{p}$
present are the inverse of averaged carrier scattering rates $\left\langle \tau_{p}^{-1}\right\rangle $.
The method of carrier lifetime calculations is given in Appendix B.
$\langle\rangle$ means taking average around chemical potential $\mu$.
For a state-resolved quantity $A_{kn}$, its average is defined as
$\left\langle A\right\rangle =\sum_{kn}A_{kn}\left[f^{\mathrm{eq}}\right]'\left(\epsilon_{kn}\right)/\sum_{kn}\left[f^{\mathrm{eq}}\right]'\left(\epsilon_{kn}\right)$,
where $\left[f^{\mathrm{eq}}\right]'$ is the derivative of Fermi-Dirac
function.\label{fig:compare_spin_carrier_relaxation}}
\end{figure}

We further examine the contributions of different scattering mechanisms
to carrier and spin lifetime respectively, as a function of temperature.
Different from spin lifetime obtained from real-time DM dynamics including
all scattering processes simultaneously, the carrier lifetime ($\overline{\tau}_{p}$)
is defined through the inverse of the averaged carrier scattering
rate($\tau_{p}^{-1}$): $\overline{\tau}_{p}=1/\left\langle \tau_{p}^{-1}\right\rangle $.
Various scattering processes (e-e, e-i and e-ph) contribute to the
total carrier scattering rates through $\tau_{p}^{-1}=(\tau_{p}\super{e-e})^{-1}+(\tau_{p}\super{e-i})^{-1}+(\tau_{p}\super{e-ph})^{-1}$.
$\langle\rangle$ means taking average around chemical potential $\mu$.
For a state-resolved quantity $A_{kn}$, its average is defined as
$\left\langle A\right\rangle =\sum_{kn}A_{kn}\left[f^{\mathrm{eq}}\right]'\left(\epsilon_{kn}\right)/\sum_{kn}\left[f^{\mathrm{eq}}\right]'\left(\epsilon_{kn}\right)$,
where $\left[f^{\mathrm{eq}}\right]'$ is the derivative of Fermi-Dirac
function. For both carrier and spin lifetime, the lifetime due to
the most dominant scattering channel is the closest to the one including
all processes (black squares in both Fig.~\ref{fig:compare_spin_carrier_relaxation}a
and b). For spin relaxation in Fig.~\ref{fig:compare_spin_carrier_relaxation}a,
at low temperature below 50 K, e-e scattering is the most dominant
process as discussed above. However, the e-ph process becomes more
dominant above 100K. On the other hand, for carrier relaxation in
Fig.~\ref{fig:compare_spin_carrier_relaxation}b, the e-i process
is dominant over a wide temperature range from low to right below
room temperature. At room temperature, for both spin and carrier lifetimes,
the e-ph scattering is the most important process (closest to the
total lifetime with all scattering processes).

Our observations differ from those in Refs. \citenum{kamra2011role}
and \citenum{marchetti2014spin}, where the authors also found that
e-e scattering dominates spin relaxation at lower temperatures, e.g.
77 K, but their results showed that at room temperature e-e scattering
can be more important than other scatterings and enhances $\tau_{s,z}$
of $n$-GaAs by about 100$\%$ with moderate doping concentrations.
The overestimate of the effects of e-e scattering at room temperature
is most likely a limitation of the semiclassical method employed therein.

Similarly, we also find that different phonon modes can play different
roles in carrier and spin relaxations as shown in Fig. \ref{fig:compare_spin_carrier_relaxation}c
and \ref{fig:compare_spin_carrier_relaxation}d. For example, at room
temperature, LO (longitudinal optical) mode is most important for
carrier relaxation but seems less important than TA (transverse acoustic)
modes for spin relaxation. The situation is the opposite at 100 K
where TA/LO is most important for carrier/spin relaxation. Our finding
that TA modes are slightly more important than LO mode in spin relaxation
at room temperature is different from what have been believed in previous
model studies\citep{marchetti2014spin,jiang2009electron}, where they
declared that the electron-LO-phonon scattering dominates spin relaxation
at high temperatures especially at room temperature. This disparity
is most likely due to differences in the e-ph matrix elements and
electronic quantities, where we used fully first-principles approaches
instead of parameterized models in previous work.

In addition, we find the total spin lifetime is the longest when considering
all scattering processes in Fig.~\ref{fig:compare_spin_carrier_relaxation}a;
in contrast, the carrier lifetime is the shortest including all scattering
mechanism in Fig.~\ref{fig:compare_spin_carrier_relaxation}b. This
follows the inverse relation between spin and carrier lifetime in
the empirical D'yakonov--Perel' (DP) mechanism\citep{dyakonov1972spin,vzutic2004spintronics}
for systems without inversion symmetry, as will be discussed in more
details in next section.

\subsubsection{Doping-level-dependence of spin lifetime and its dominant relaxation
mechanism\label{subsec:doping}}

\begin{figure*}
\includegraphics[scale=0.24]{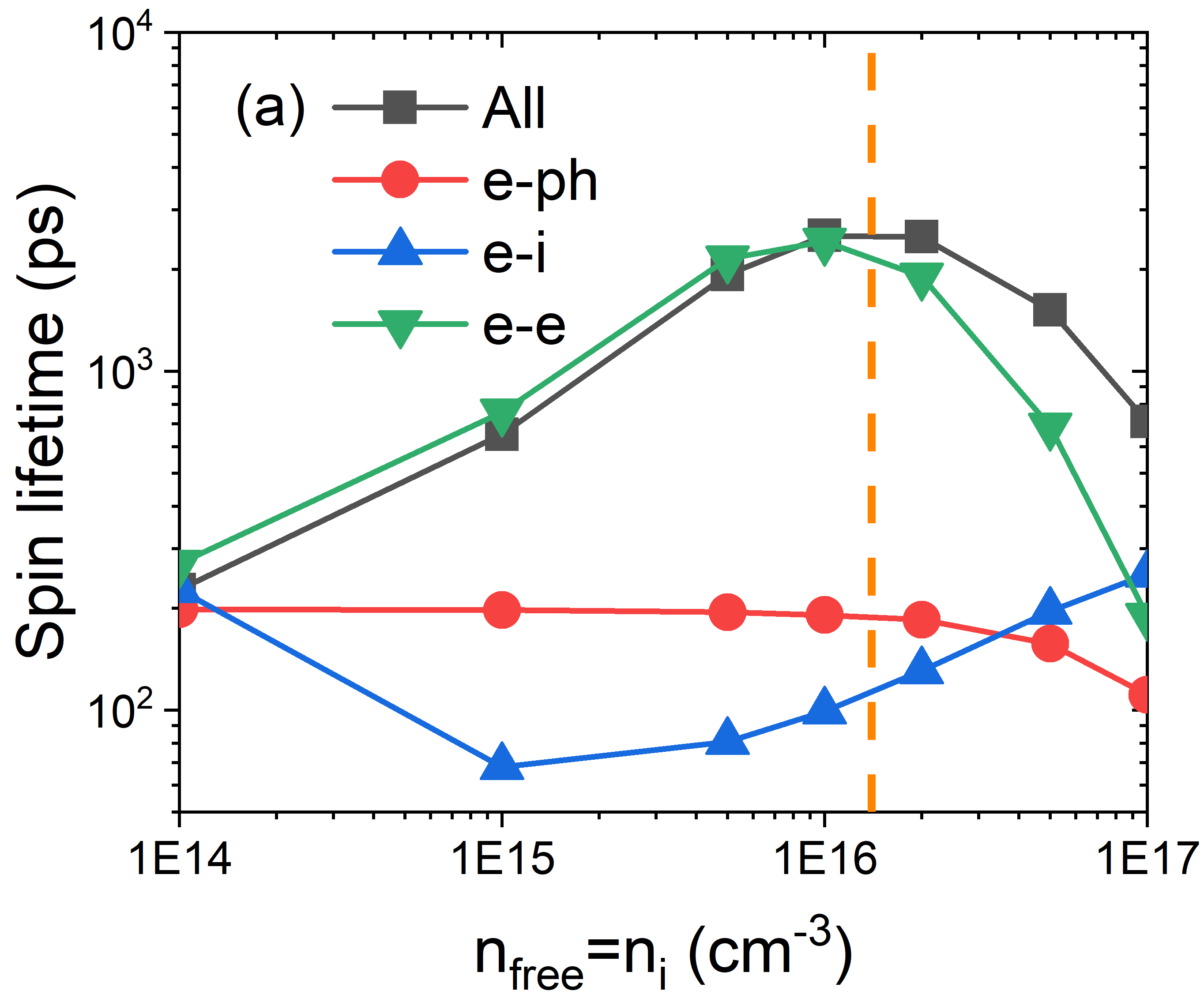} \includegraphics[scale=0.24]{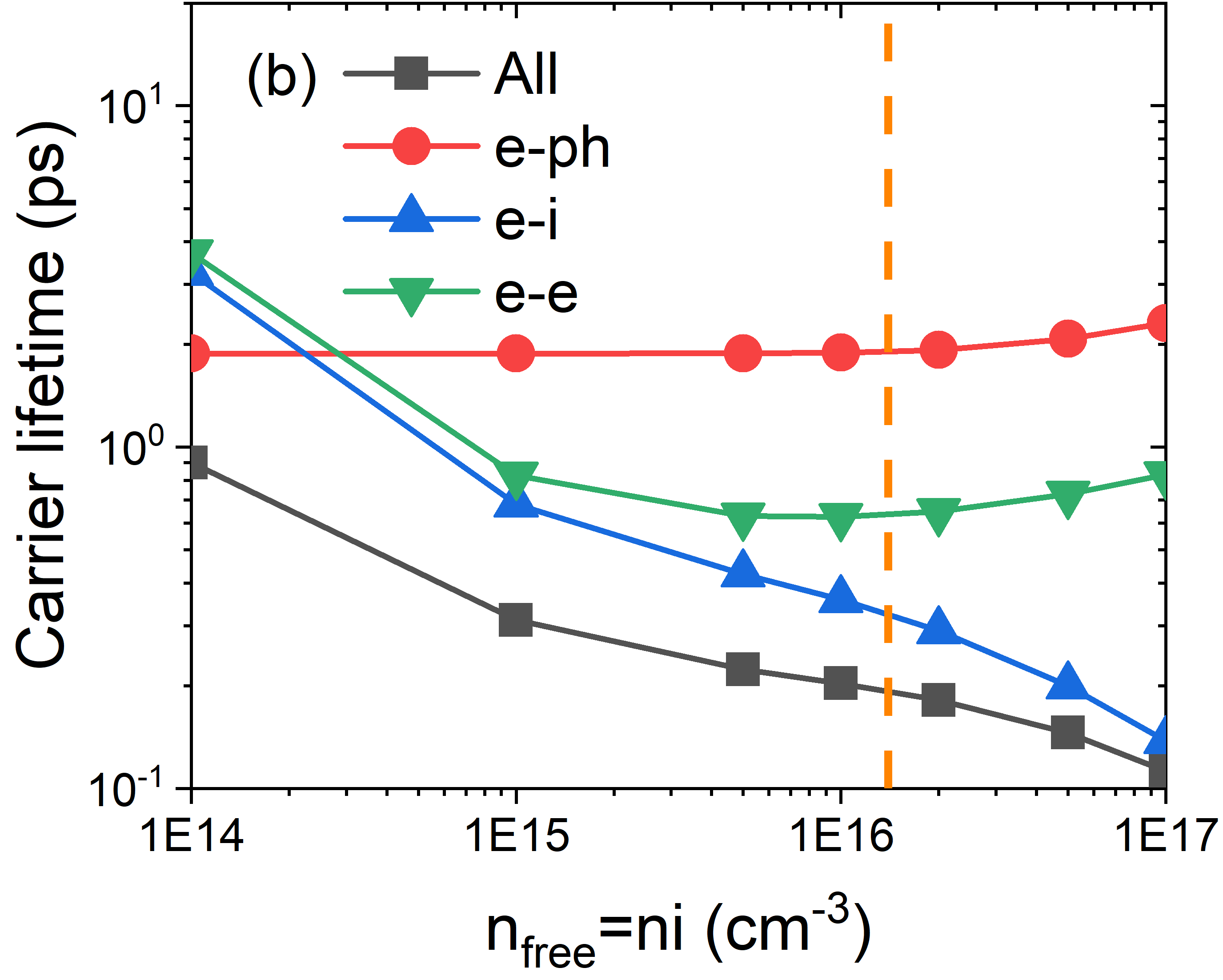}
\includegraphics[scale=0.24]{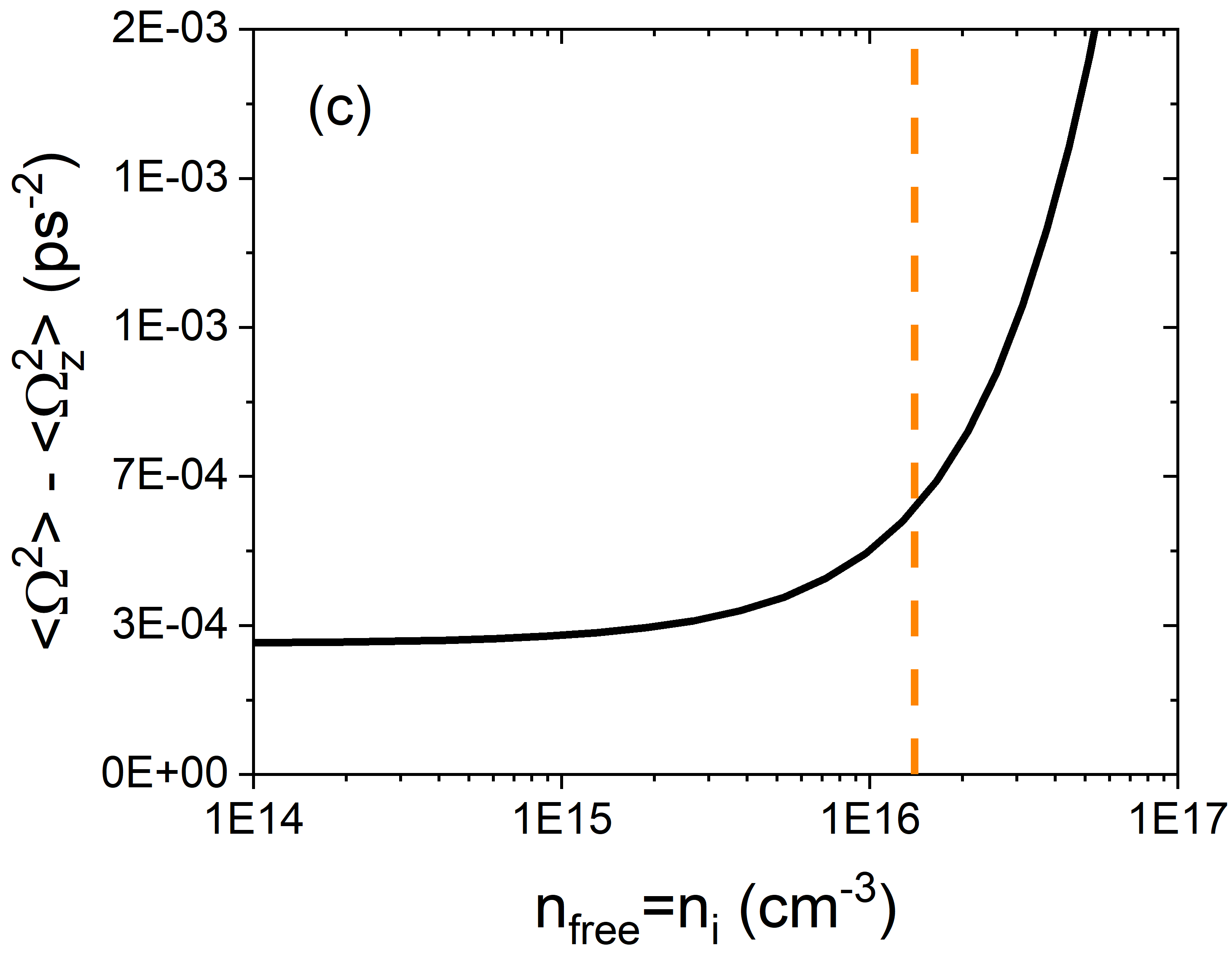} \caption{(a) Spin and (b) carrier lifetimes of $n$-GaAs with different doping
concentrations at 30 K with different scattering mechanisms. ``All''
represents all the e-ph, e-i and e-e scattering mechanisms being considered.
(c) $\left\langle \vec{\Omega}^{2}\right\rangle -\left\langle \Omega_{i}^{2}\right\rangle $
as a function of carrier density, where $\vec{\Omega}$ is the Larmor
frequency due to the ``internal'' magnetic field computed from first-principles,
which describes the SOC term induced by inversion asymmetry.\label{fig:doping}}
\end{figure*}

Figure~\ref{fig:doping} shows the carrier and spin lifetimes with
different doping density $n_{i}$ at 30 K with individual and total
scattering pathways, respectively. Similar to temperature dependence
and phonon contributions, it is also found that the roles of different
scattering mechanism differ considerably between spin and carrier
relaxation processes. Specifically, for the carrier relaxation in
Fig.~\ref{fig:doping}b, except when $n_{i}$ is very low (e.g. at
$10^{14}$ cm$^{-3}$), the electron-impurity scattering (e-i) dominates,
similar to the case of carrier lifetime over a large range of temperature
at a moderate doping in Fig.~\ref{fig:compare_spin_carrier_relaxation}b.
On the other hand, for the spin relaxation in Fig.~\ref{fig:doping}a,
the e-e scattering dominates except at very high concentration (above
$10^{17}$ cm$^{-3}$), while e-i scattering is only important in
the very high doping region (close to or above $10^{17}$ cm$^{-3}$).

Figure~\ref{fig:doping} shows the calculated $\tau_{s}$ has a maximum
at $n_{i}\,=\,1\text{-}2\times10^{16}$ cm$^{-3}$, and $\tau_{s}$
decreases fast with $n_{i}$ going away from its peak position. This
is in good agreement with the experimental finding in Ref. \citenum{kikkawa1998resonant},
which also reported $\tau_{s}$ at $n_{i}=10^{16}$ cm$^{-3}$ is
longer than $\tau_{s}$ at other lower and higher $n_{i}$ at a low
temperature (a few Kelvin). The $n_{i}$ dependence of $\tau_{s}$
may be qualitatively interpreted from the commonly used empirical
DP relation\citep{vzutic2004spintronics} for inversion-asymmetric
systems, $\tau_{s,i}\sim\tau_{s,i}\super{DP}=1/\left[\overline{\tau}_{p}\cdot\left(\left\langle \vec{\Omega}^{2}\right\rangle -\left\langle \Omega_{i}^{2}\right\rangle \right)\right]$,
where $\overline{\tau}_{p}$ is the carrier lifetime, and $\vec{\Omega}$
is the Larmor frequency due to the ``internal'' magnetic field,
which describes the SOC term induced by inversion asymmetry. For spin
1/2 systems, the internal magnetic field at ${\bf k}$ ($\vec{\Omega}_{{\bf k}}$)
will induce an energy splitting $\Delta_{{\bf k}}$ and polarize the
spin along the direction of $\vec{\Omega}_{{\bf k}}$. Previously,
$\vec{\Omega}_{{\bf k}}$ was mostly obtained with model Hamiltonian
with Dresselhaus SOC field\citep{dresselhaus1955spin}, which is rather
qualitative. Instead, we obtained k-dependent internal magnetic field
$\vec{\Omega}_{{\bf k}}$ from first-principles calculations, by using
$\Omega_{{\bf k},i}=2\Delta_{{\bf k}}\cdot s_{{\bf k},i}\super{exp}/\hbar$,
where $s_{{\bf k},i}\super{exp}$ is the spin expectation value.

From Fig.~\ref{fig:doping}, we find that with $n_{i}$ from $10^{14}$
cm$^{-3}$ to $5\times10^{15}$ cm$^{-3}$, carrier lifetime $\overline{\tau}_{p}$
decreases rapidly (black curve in Fig.~\ref{fig:doping}b) and $\left\langle \vec{\Omega}^{2}\right\rangle -\left\langle \Omega_{i}^{2}\right\rangle $
remains flat in Fig.~\ref{fig:doping}c, which may explain why spin
lifetime ($\tau_{s}$) increases in Fig.~\ref{fig:doping}a based
on the DP relation; however, when $n_{i}>10^{16}$ cm$^{-3}$, $\overline{\tau}_{p}$
decreases with a similar speed but $\left\langle \vec{\Omega}^{2}\right\rangle -\left\langle \Omega_{i}^{2}\right\rangle $
experiences a sharp increase, which may explain why spin lifetime
decreases in Fig.~\ref{fig:doping}b and owns a maximum at $10^{16}$
cm$^{-3}$.

Note that although the above empirical DP relation is intuitive to
understand the cause of doping-level dependence of spin lifetime,
it may break down when we evaluate individual scattering processes.
For example, when $n_{i}$ increases from $10^{14}$ cm$^{-3}$ to
$10^{15}$ cm$^{-3}$, both carrier lifetime $\overline{\tau}_{p}$
and spin lifetime $\tau_{s,z}$ due to e-i scattering decrease while
the internal magnetic field remains unchanged. Moreover, the simple
empirical relation cannot possibly explain our first-principles results
that the e-e and e-i scatterings have largely different contributions
in carrier and spin relaxation. First-principles calculations are
critical to provide unbiased mechanistic insights to spin and carrier
relaxation of general systems.

\subsection{Applications to few-layer WSe$_{2}$}

\subsubsection{Spin/valley relaxation of resident holes of monolayer WSe$_{2}$}

For holes of monolayer WSe$_2$, spin/valley relaxation is mostly
determined by intervalley spin-flip scattering processes between $K$ and $K'$
valleys because of the spin-valley locking. Previously, we reported spin/valley
lifetimes of resident holes of monolayer TMDs at T$\geq$50 K with
e-ph scattering\cite{xu2020spin}. At very low temperatures, e.g.,
10 K, intervalley e-ph scattering is however not activated as the
corresponding phonon occupation is negligible; therefore, other scattering
mechanisms are necessary to be included. Note that e-e scattering should not play
an important role in spin relaxation of holes of TMDs. The reason
is: The e-e scattering is a two-particle process where a transition
is accompanied by another transition with energy and momentum being
conserved. Considering the fact that only the highest occupied band
is involved (see band structure in Fig. S5) in dynamics of TMD holes, for an e-e process, a $K$$\rightarrow$$K'$
($K'$$\rightarrow$$K$) spin-flip transition must be accompanied
by an opposite $K'$$\rightarrow$$K$ ($K$$\rightarrow$$K'$) spin-flip
transition. Overall, e-e scattering processes have negligible contributions to spin relaxation
of TMD semiconductors. As a result, we will include only e-ph and e-i scatterings for WSe$_2$.
We use the supercell method to compute e-i scattering matrix elements
for neutral defects with self-consistent SOC and more details can be found in Appendix A.

\begin{figure}
\includegraphics[scale=0.15]{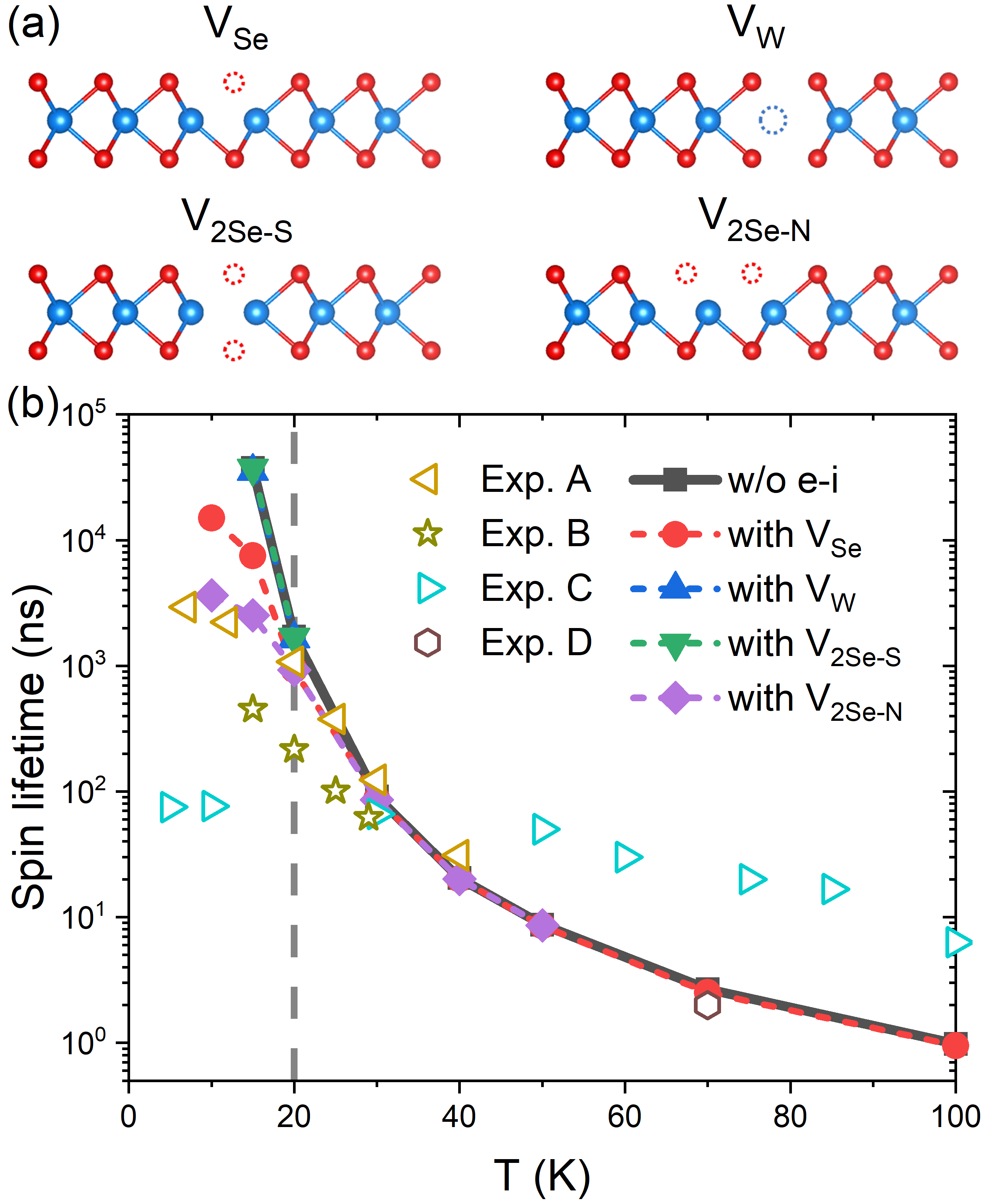}

\caption{
(a) The schematics of four types of impurities in WSe$_{2}$. (b)
Spin lifetimes of holes of monolayer WSe$_{2}$ with a relatively
low hole density $10^{11}$ cm$^{-2}$ with impurities compared with
experimental data. Exp. A, B, C and D are experimental data from Refs. \citenum{li2021valley}, \citenum{goryca2019detection}, \citenum{song2016long} and \citenum{yan2017long}, respectively.
The choices of impurity concentration $n_{i}$ of different impurities
are given in the main text.\label{fig:monolayer}}
\end{figure}

Experimentally several types of impurities/defects exist in TMD
samples. Here we pick four types of impurities with different symmetries
and chemical bonds (see Fig. \ref{fig:monolayer}(a)) - Se vacancy
($\mathrm{V_{Se}}$), two neighboring Se vacancies ($\mathrm{V_{2Se-N}}$), W vacancy ($\mathrm{V_{W}}$) and two Se vacancies with the same in-plane position ($\mathrm{V_{2Se-S}}$). As most point defects are relatively deep with large ionization energies in semiconducting TMDs\citep{wang2019excitation}, we mostly consider neutral defects here. According to Refs. \citenum{edelberg2019approaching,rhodes2019disorder,yankowitz2015local}, the impurity concentration 
$n_{i}$ ranges from 8$\times$10$^{10}$ to 10$^{14}$ cm$^{-2}$
depending on samples. Considering that $\mathrm{V_{Se}}$ is often regarded as
the most abundant impurity, we choose a reasonable impurity density $n_{i}$ of $\mathrm{V_{Se}}$ - 7$\times$10$^{11}$ cm$^{-2}$, within the experimental range and for better comparison with experimental $\tau_{s}$ at T$\geq$20 K shown in Fig. \ref{fig:monolayer}(b). 
$n_{i}$ of $\mathrm{V_{2Se-N}}$ is chosen
as 8$\times$10$^{9}$ cm$^{-2}$, two order of magnitude lower than $\mathrm{V_{Se}}$
because of its larger formation energy\cite{yankowitz2015local} and 
better comparison with experimental $\tau_{s}$. $n_{i}$ of $\mathrm{V_{W}}$
and $\mathrm{V_{2Se-S}}$  are chosen arbitrarily as we find they have rather
weak effects on spin relaxation and are 7$\times$10$^{11}$ and 3.5$\times$10$^{11}$
cm$^{-2}$, respectively.

\begin{figure}
\includegraphics[scale=0.146]{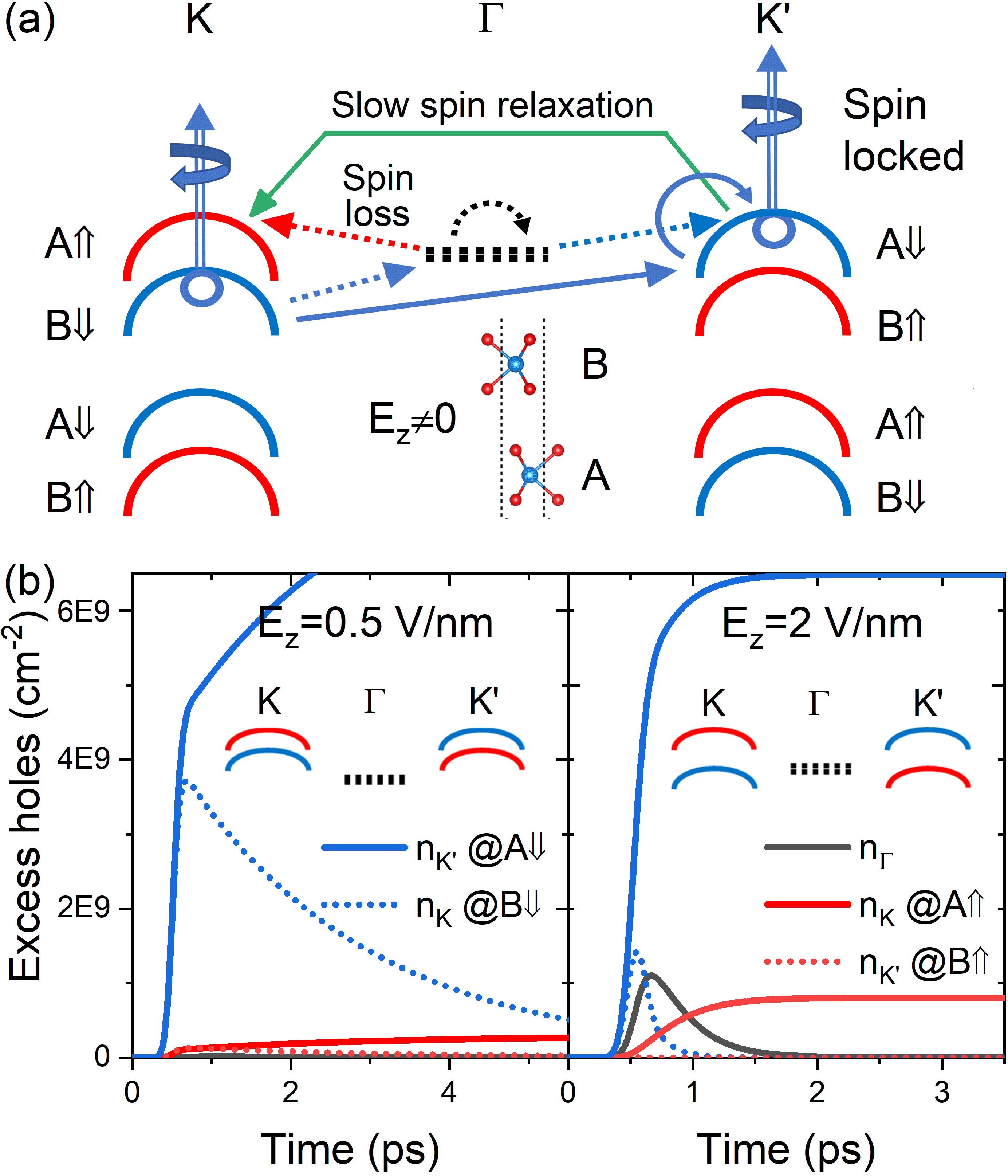}

\caption{
(a) Schematic diagram of scattering pathways of excited holes in 1$\%$
compressed bilayer WSe$_{2}$ with a low equilibrium hole density
and 1.4$\times$10$^{12}$ cm$^{-2}$ $\mathrm{V_{Se}}$ (double of the monolayer value in Fig.~\ref{fig:monolayer}) under finite $E_{z}$ during and after a circularly polarized pump, 
which initially excite holes (labelled as half circles) 
at both $K$ and $K'$ valleys and two top valence bands (``A$\Downarrow$'' at $K'$ and ``B$\Downarrow$'' at $K$).
The lower-energy excited holes will decay to the band edge
(most to ``A$\Downarrow$'' at $K'$ and fewer to ``A$\Uparrow$'' at $K$)
through different scattering pathways.
A state being labeled by ``A''(``B'') means the wavefunction
of this state is mostly localized in layer A(B) of the bilayer. 
$\Uparrow$ and $\Downarrow$ represent spin-up and spin-down, respectively.
The color of electronic state represents spin polarization - red and blue
mean spin-up and spin-down, respectively. 
(b) Time evolutions of valley-
and band-resolved (excited or excess) hole densities at 50 K under
two $E_{z}$ with a circularly polarized pump. The pump energy is
selected to excite electrons at two top valence bands. The pump center
is at 0.5 ps. $n_{V}$ represents excess hole density at valley $V$.
The insets of panel (b) are the schematics of energies of two top
valence bands at $K$, $K'$ and $\Gamma$ valleys under two $E_{z}$ (see calculated band structures in SI Fig.~S7).
\label{fig:bilayer_dynamics}}
\end{figure}

From Fig. \ref{fig:monolayer}, we first find that 
at T\textgreater 20 K, spin relaxation is almost driven
by e-ph scattering and impurities can only affect spin relaxation
at T$\leq$20 K. For the effects of different impurities on spin relaxation,
we have $\mathrm{V_{2Se-N}}$ $\gg$ $\mathrm{V_{Se}}$ $\gg$ $\mathrm{V_{W}}$ $\sim$ $\mathrm{V_{2Se-S}}$.
Such differences are directly related to the large differences among various impurities in 
electron-impurity matrix elements for
the intervalley processes (scattering between $K$ and $K'$ valley), i.e. much larger matrix elements $|g_i|$ for intervalley scattering at $\mathrm{V_{2Se-N}}$ and $\mathrm{V_{Se}}$ compared to the ones at $\mathrm{V_{W}}$ and $\mathrm{V_{2Se-S}}$
as shown in SI Fig.~S9.
Moreover, the temperature dependence of $\tau_{s}$ with $\mathrm{V_{2Se-N}}$
is much weaker and in better agreement with experiments than that
with $\mathrm{V_{Se}}$. Therefore, the observed weak temperature dependence
in some experiments is probably related to the existence of larger size impurities
with lower symmetries (e.g. $\mathrm{V_{2Se-N}}$). Our observations suggest that the local symmetry
and chemical bonds surrounding an impurity have large impact on spin relaxation.

Additionally, we have simulated spin lifetimes of
monolayer WSe$_{2}$ at 15 K with different hole densities and find
a strong hole density dependence. The related results are shown in SI Fig. S12.

\subsubsection{Ultrafast dynamics of holes of bilayer WSe$_{2}$}

Understanding detailed dynamical processes and related scattering mechanism can help develop strategy of controlling and manipulating spin/valley relaxation through tuning external fields, materials composition, and strain. In the following, we will first identify the scattering pathways of excited holes and spins in bilayer WSe$_2$ (AB stacking as shown in Fig.~\ref{fig:bilayer_dynamics}a).
Next through real-time simulations, we determine dynamical quantities like spin lifetimes and valley polarization at different external fields and strain. 
Finally, we show the carrier occupation on each layer is fully spin polarized and can be switched by an external electric field.

According to previous studies\cite{kim2021thickness,zhao2013origin} and our calculations, for valence bands
of unstrained bilayer WSe$_{2}$, $\Gamma$ valley is slightly higher
than $K$/$K'$ valley, which is usually undesirable. 
The $K$/$K'$ valley can be pushed higher than $\Gamma$ valley by applying slight
in-plane compressive strain (corresponding band structures can be found in SI Fig.~S6). 
Moreover, under zero $E_{z}$, the bilayer WSe$_2$ has inversion symmetry, leading to carriers being equally populated at $K$ and $K'$ valleys and at two layers, with spin up and down degeneracy. 
A finite $E_z$ can break inversion symmetry (thus break Kramers degeneracy) and induce non-zero layer polarization or layer population difference. For example, two top valence bands from layer A with $\Uparrow$ (up spin) and layer B with $\Downarrow$ (down spin) are degenerate at $K$ without electric field but split under electric field. 
Thus each band is associated with a particular spin channel, valley and layer, i.e.  spin-valley-layer locking effect. 
Also by tuning the sign and magnitude of $E_z$,
we are able to control various physical quantities like layer pseudospin, band splitting energy, etc.
Therefore, to ensure spin-valley-layer locking effects
being observed, we will study slightly compressed inversion-symmetric
bilayer WSe$_{2}$ under finite $E_{z}$.

Figure \ref{fig:bilayer_dynamics}(a) shows the scattering pathway schematics for hole bands (half circles) during the first few ps of slightly compressed (1$\%$) bilayer WSe$_{2}$ excited by a circularly polarized pump pulse under finite $E_{z}$ at
50 K, before exciton recombination processes happen typically at tens of ps timescale at this temperature. \cite{palummo2015exciton,goodman2017exciton}
Similar to the GaAs case (see Sec. \ref{subsec:GaAs_ultrafast}), the
hole spins will undergo the following processes: optical generation, decay to band edges and at the end slow relaxation.
Initially, holes with the same spin polarization are excited at
both $K$ and $K'$ valleys and two layers equally (e.g. down spin holes generated at $K$ valley and layer B and $K'$ valley and layer A as shown in Fig.~\ref{fig:bilayer_dynamics}(a)). During and after
the excitations, lower-energy holes at $K$ ($K'$) valley will decay
to the band edge through two possible
scattering pathways: (i) Direct pathway through interlayer spin-conserving
scattering (solid blue line); (ii) Indirect pathway through $\Gamma$-valley-related
scattering (dashed lines for both spin conserving and flip processes). After all holes decay to the band edge, 
most of their carried spins are ``locked'' at a certain layer and valley (e.g., ``A$\Downarrow$'' at $K'$ in Fig.~\ref{fig:bilayer_dynamics}) due to weak intervalley spin-flip scattering (green arrow in Fig.~\ref{fig:bilayer_dynamics}(a)), which
is the so-called ``spin-valley-layer locking''.

\begin{figure}
\includegraphics[scale=0.13]{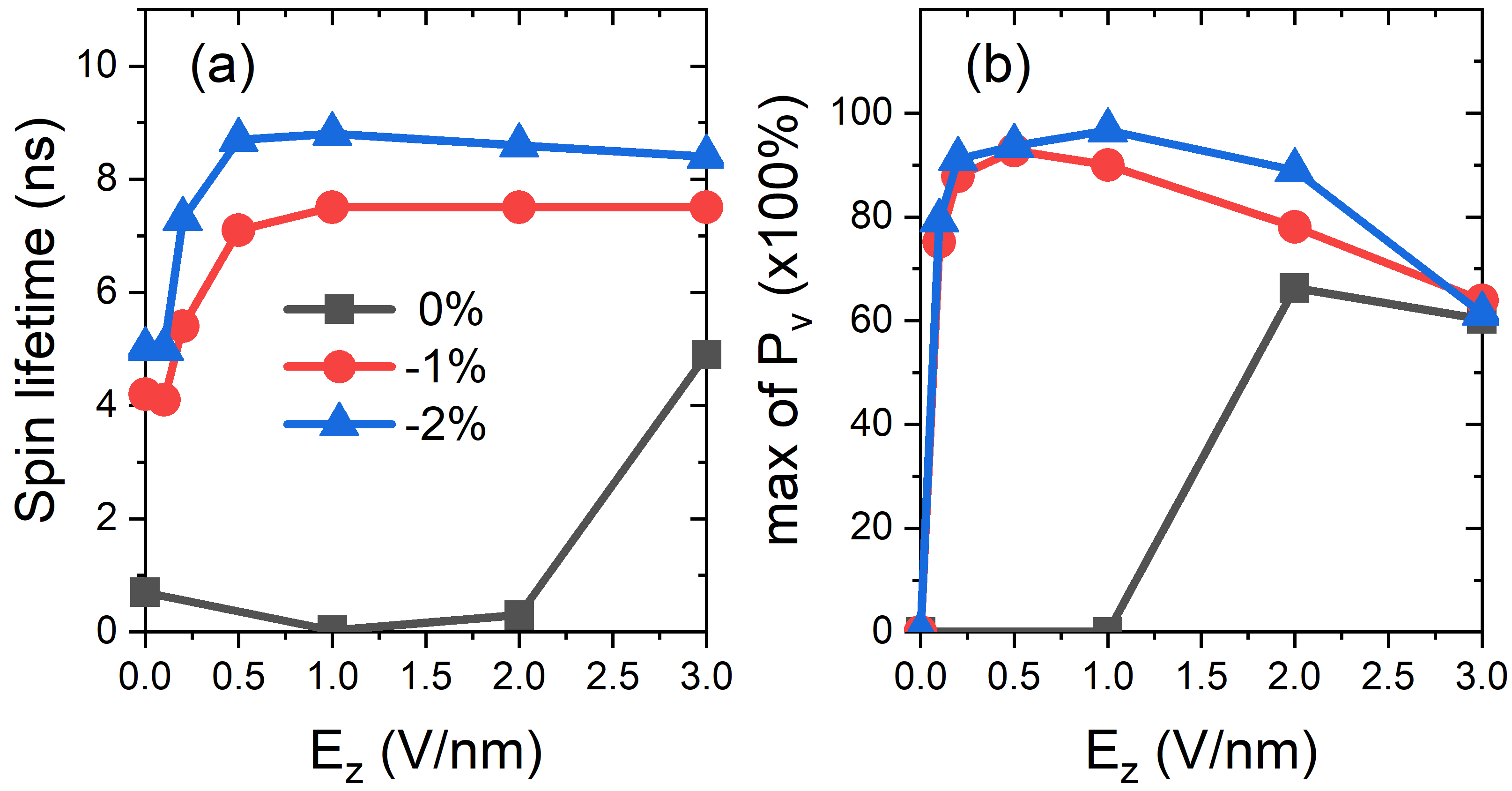}
\caption{
Spin lifetime and maximum valley polarization $P_{V}=|n_{K}-n_{K'}|/n_{tot}$
of bilayer WSe$_{2}$ at 50 K, where $n_{K(K')}$ is excess hole density
at $K\left(K'\right)$ valley and $n_{tot}$ is total excess hole density. 
Negative strain means compressive strain.
$P_{V}$ will be 90$\%$ of its maximum shortly after the pump and
reach the maximum at a time from 1 to 20 ps depending on $E_z$ and the strain.
After reaching its maximum, since most carriers/spins have already decayed to the band edge, $P_{V}$ can relax only very slowly through intervalley spin-flip scattering.
This implies that having a high maximum of $P_{V}$ is important 
to ensure a high $P_{V}$ during a long time.
}\label{fig:spin_lifetime_and_valley_pol}
\end{figure}

In Fig. \ref{fig:bilayer_dynamics}(b), we show time evolutions of
valley- and band-resolved (excited or excess) hole densities under
two $E_{z}$. It can be seen that under a low $E_{z}$ (0.5 V/nm, Fig. \ref{fig:bilayer_dynamics}(b) left panel),
the main scattering pathway is the direct one mentioned above, i.e. the spin down holes scattered from B $\Downarrow$ at $K$ in dashed blue line with decreasing population  to A $\Downarrow$ at $K'$ in solid blue line with increasing population. Under
a higher $E_{z}$ (2 V/nm, Fig. \ref{fig:bilayer_dynamics}(b) right panel), although the direct scattering pathway
still exists, the indirect one through the $\Gamma$ valley also becomes important because the band energy at $\Gamma$ is pushed higher than the second valence band at $K$ and $K'$ under this electric field (see inset of Fig. \ref{fig:bilayer_dynamics}(b) right panel). Here occupation at B $\Downarrow$ at $K$ in dashed blue line rapidly decreased while occupation at $\Gamma$ with both up and down spins in solid black temporarily increased through indirect scattering, 
and most importantly the A $\Uparrow$ at $K$ in solid red also increased due to the scattering through $\Gamma$ valley. Increased population at A $\Uparrow$ at $K$ represents weakening the spin-valley-locking effect. Therefore, the indirect scattering pathway will lead to the reduction of spin density and valley polarization.

\begin{figure}
\includegraphics[scale=0.35]{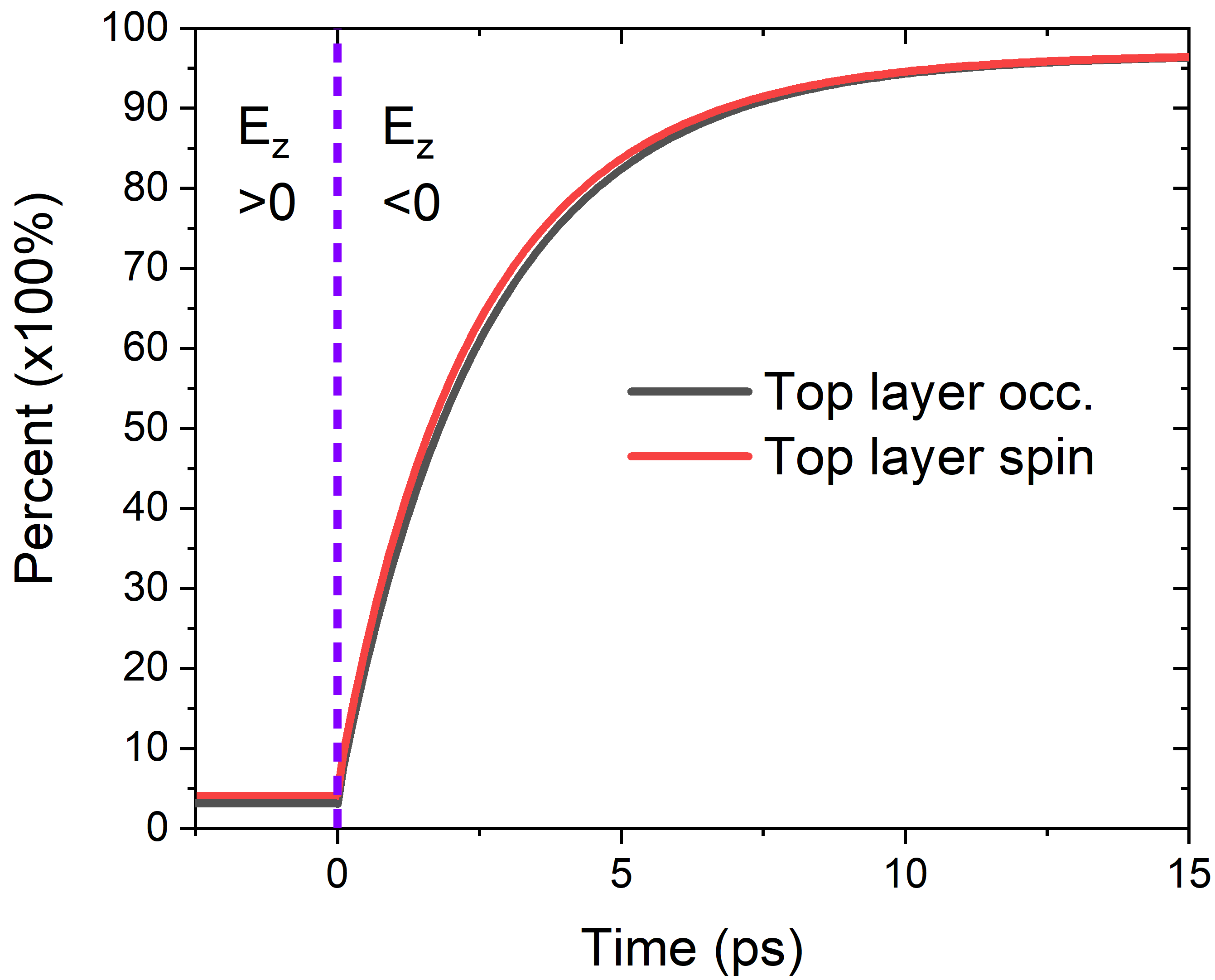}

\caption{
Time evolution of top layer hole occupation $f_{h}^{\mathrm{top}}$
and spin $s_{z}^{\mathrm{top}}$ (see their definitions in Appendix D) 
normalized by corresponding total quantities of compressed bilayer WSe$_{2}$ at 50 K after the sign
of $E_{z}$ is switched at $t=0$. At $t=-10$ ps, a pump pulse centered
at $t=-9.5$ ps with $\tau_{\mathrm{pump}}=100$ fs is started to
be applied and real-time density matrix dynamics is run under $E_{z}=0.5$
V/nm until $t=0$ to allow holes/spins to decay to the band edge where
states are localized in bottom layer. When the sign of $E_{z}$ is
suddenly switched, holes/spins are still localized in the same bottom
layer but the new eigenstates around the band edge are localized in
top layer, thus holes/spins will transfer from bottom to top layer.
\label{fig:switching}
}
\end{figure}

We then show two key dynamical quantities - spin lifetime $\tau_{s}$ in Fig.~\ref{fig:spin_lifetime_and_valley_pol}(a) and maximum
valley polarization of bilayer WSe$_{2}$ in Fig.~\ref{fig:spin_lifetime_and_valley_pol}(b) as a function of $E_{z}$
with different percentages of strain. Valley polarization is defined as $P_{V}=|n_{K}-n_{K'}|/n_{tot}$,
where $n_{K(K')}$ is excess hole density at $K\left(K'\right)$ valley
and $n_{tot}$ is total excess hole density. 
A high maximum of $P_{V}$ is necessary to ensure a high $P_{V}$ for extended time.
This is because: when $P_{V}$ reaches its maximum, since most carriers/spins have already decayed to the band edge at the same time, 
$P_{V}$ will decay very slowly through intervalley spin-flip scattering.
Obviously, unstrained system is not suitable to utilize spin-valley-layer locking considering
its short spin lifetimes and low maximum $P_{V}$ under a relatively low electric field. As we discussed
above, a slight strain is helpful to ensure $K/K'$ valleys are sufficiently
higher than $\Gamma$ valley. Moreover, from Fig. \ref{fig:spin_lifetime_and_valley_pol}, an optimized range of $E_{z}$ is from $\sim$0.2 to $\sim$2
V/nm, where the system shows long $\tau_{s}$ and high maximum $P_{V}$, consistent
with the fact mentioned above: indirect scattering pathway, which becomes
important under high $E_{z}$, will cause loss of spin and
valley polarization.

Finally, in a bilayer, the percentage of carriers/spins localized in
top/bottom layer at equilibrium can be controlled by a static $E_{z}$.
For device applications, it may be desirable to dynamically and spatially
tune the locations of holes/spins by controlling $E_{z}$. For this
purpose, a knowledge of the speed of carriers/spins transferring between
two layers can be useful.

To extract the time scale of such transfer, we first generate
holes/spins by applying a circularly polarized pump pulse centered
at $-9.5$ ps and let the system evolves until $t=0$ under $E_{z}=0.5$
V/nm to ensure almost all holes/spins being localized in bottom layer,
then by switching $E_{z}$ suddenly to $-0.5$ V/nm at $t=0$, holes/spins
will start to transfer to top layer.

In Fig. \ref{fig:switching}, we show time evolution of
top layer hole occupation $f_{h}^{\mathrm{top}}$ and spin $s_{z}^{\mathrm{top}}$
(see their definitions in Appendix D)
normalized by corresponding total quantities of 2\% compressed bilayer
after the sign of $E_{z}$ is switched.
From Fig. \ref{fig:switching}, at 50 K, 90\% switching of $f_{h}^{\mathrm{top}}$
and $s_{z}^{\mathrm{top}}$ takes $\sim$6 ps. This time constant
is much shorter than $\tau_{s}$, which means such tuning is fast enough to use an electric field as a "switch" in spintronic devices.

\section{Conclusions}

In this article, we present a first-principles real-time density-matrix
approach to simulate ultrafast spin-orbit-mediated spin dynamics in
solids with arbitrary crystal symmetry. The complete \emph{ab initio}
descriptions of pump, probe and three scattering processes - the electron-phonon,
electron-impurity and electron-electron scattering in the density-matrix
master equation, allows us to directly simulate the nonequilibrium
ultrafast pump-probe measurements and makes our method applicable
to any temperatures and doping levels. This method has been applied
to simulate spin relaxation of $n$-GaAs. We confirm that relaxation
time of Kerr rotation and that of spin observables are almost identical
and find that relaxation time of spin polarization is relatively robust,
i.e. insensitive to how spin imbalance is initialized. Furthermore,
we have studied the temperature and doping-level dependencies of spin
lifetime and examined the roles of various scattering mechanisms.
Overall our theoretical results are in good agreement with experiments.
Importantly, our first-principles simulations provide rich mechanistic
insights of spin relaxation of $n$-GaAs: we point out that although
at low temperatures and moderate doping concentrations e-i scattering
dominates carrier relaxation, e-e scattering is the most dominant
process in spin relaxation. The relative contributions of phonon modes
also vary considerably between spin and carrier relaxation. We have
further examined ultrafast dynamics in few-layer WSe$_{2}$ with
realistic impurities. We find that spin relaxation can highly depend on local symmetry and chemical bonds
surrounding impurities. For the bilayer, we identify the scattering
pathways of holes in ultrafast dynamics and determine relevant dynamical
properties, including $\tau_{s}$, maximum valley polarization and
layer population/spin switch time, which are essential to utilize
its unique spin-valley-layer locking effects. Our method opens up
the pathway to predict spin relaxation and decoherence for general
materials and provide unbiased insights and guidelines to experimental
materials design, which have the potential to revolutionize the field
of spintronics and quantum information technologies.

\section*{Acknowledgements}

We thank Hiroyuki Takenaka for helpful discussions. This work is supported
by National Science Foundation under grant No. DMR-1956015 and the Air Force Office of Scientific Research under AFOSR Award No. FA9550-YR-1-XYZQ. A. H.
acknowledges support from the American Association of University Women(AAUW)
fellowship program. This research used resources of the Center for
Functional Nanomaterials, which is a US DOE Office of Science Facility,
and the Scientific Data and Computing center, a component of the Computational
Science Initiative, at Brookhaven National Laboratory under Contract
No. DE-SC0012704, the lux supercomputer at UC Santa Cruz, funded by
NSF MRI grant AST 1828315, the National Energy Research Scientific
Computing Center (NERSC) a U.S. Department of Energy Office of Science
User Facility operated under Contract No. DE-AC02-05CH11231, the Extreme
Science and Engineering Discovery Environment (XSEDE) which is supported
by National Science Foundation Grant No. ACI-1548562 \citep{xsede},
and resources at the Center for Computational Innovations at Rensselaer
Polytechnic Institute.

\section*{Appendix A: Interaction Hamiltonian terms and matrix elements}

Three interaction Hamiltonian terms in Eq. \ref{eq:H} read

\begin{align}
H\sub{e\text{-}ph}= & \sum_{12q\lambda}c_{1}^{\dagger}c_{2}\left(g_{12}^{q\lambda-}b_{q\lambda}+g_{12}^{q\lambda+}b_{q\lambda}^{\dagger}\right),\\
H\sub{e\text{-}i}= & n_{i}V\sub{cell}\sum_{12}c_{1}^{\dagger}c_{2}g_{12}^{i},\\
H\sub{e\text{-}e}= & \sum_{1234}c_{1}^{\dagger}c_{2}^{\dagger}c_{3}c_{4}g_{1234}\super{e\text{-}e}.
\end{align}

The e-ph matrix $g^{q\lambda\pm}$ are computed with self-consistent
spin-orbit coupling and Wannier interpolation by the supercell method.

Here we assume impurity density is sufficiently low 
and the average distance between neighboring impurities is sufficiently long 
so that the interactions between impurities are negligible. 
The e-i matrix $g^{i}$ is
\begin{align}
g_{13}^{i}= & \left\langle 1\right|\Delta V^{i}\left|3\right\rangle ,\\
\Delta V^{i}= & V^{i}-V^{0},
\end{align}
where $V^{i}$ is the potential of the impurity system
and $V^{0}$ is the potential of the pristine system. In this work, $g^{i}$ for neutral and ionized impurities
are computed differently as follows.

For neutral impurities, $V^{i}$ is computed  using a large supercell including an impurity with self-consistent SOC at DFT. This is important for including the detailed potential profile and chemical bonding environment for different impurities. 
To speed up the
supercell convergence, we used the potential alignment method developed
in Ref. \citenum{sundararaman2017first}. 
We checked the supercell
size convergence of $g^{i}$ for neutral $\mathrm{V_{Se}}$ in monolayer WSe$_{2}$
and found that the corresponding spin lifetime with 6$\times$6
and 8$\times$8 supercells differs by only a few percent. 
Thus
6$\times$6 supercell is enough for $g^{i}$ of
neutral impurities in monolayer WSe$_{2}$.

For ionized impurities, we use approximate impurity potentials as detailed below.
In general, $\Delta V^{i}$ may be separated into
two terms $\Delta V^{i}=\Delta V_{\mathrm{ns}}^{i}+\Delta V_{\mathrm{soc}}^{i},$
where $\Delta V_{\mathrm{ns}}^{i}$ is the spin-independent part and 
$\Delta V_{\mathrm{soc}}^{i}=\left[\hbar/\left(4m^{2}c^{2}\right)\right]\nabla\left(\Delta V_{\mathrm{ns}}^{i}\right)\times\vec{p}\cdot\vec{\sigma}$
is the SOC correction. For ionized impurities, we approximate $\Delta V_{\mathrm{ns}}^{i}$
as the potential of point charge and is simply the product of the
impurity charge $Z$ and the screened Coulomb potential\citep{jacoboni2010theory}, i.e., $\Delta V_{\mathrm{ns}}^{i}=ZV^{\mathrm{scr}}$.
Such approximate describes the long-range part of spin-independent differential impurity potential $\Delta V_{\mathrm{ns}}^{i}$
accurately, which is often the most important contribution from ionized impurities. Considering that $\Delta V_{\mathrm{soc}}^{i}$ is commonly
neglected in previous theoretical studies on spin relaxation\cite{tamborenea2003spin,jiang2009electron}
whenever screened Coulomb potential $V^{\mathrm{scr}}$ is used, we will not include $\Delta V_{\mathrm{soc}}^{i}$
for ionized impurities either. Note that such potential for ionized impurities still relaxes spin through spin-mixing and spin-precession.
The e-e matrix $g\super{e\text{-}e}$ is
\begin{align}
g_{1234}\super{e\text{-}e}= & \left\langle 1\left(r\right)\right|\left\langle 2\left(r'\right)\right|V\left(r-r'\right)\left|3\left(r\right)\right\rangle \left|4\left(r'\right)\right\rangle ,
\end{align}
where $V\left(r-r'\right)$ is the screened electron-electron interaction. 
The SOC corrections on $V\left(r-r'\right)$ \citep{aryasetiawan2008generalized,grimaldi1997theory} will not be included 
similar to the ionized impurity case.
Thus, $V\left(r-r'\right)$ is simply the screened Coulomb potential $V^{scr}$.
Therefore, in both calculations of $g^{i}$ and $g\super{e\text{-}e}$, the computation of the screened Coulomb potential $V^{scr}$ is of key importance.

Currently, we use the static RPA (Random Phase Approximation) dielectric
function for the screening and neglect local-field effects. We then
show the e-e self-energy ($\mathrm{Im\Sigma}$) obtained with such
dielectric function well reproduces the one obtained with dynamically
screened Coulomb interaction with full RPA dielectric matrix in the
relevant energy range as shown in Fig.~\ref{fig:ImSigma_benchmark}.
The dielectric function has the form
\begin{align}
\epsilon\left(\vec{q}\right)= & \epsilon_{s}\epsilon\super{intra}\left(\vec{q}\right),
\end{align}

where $\epsilon_{s}$ is the static background dielectric constant
and can be calculated by Density Functional Perturbation Theory (DFPT)\citep{wu2005systematic}.
$\epsilon\super{intra}\left(\vec{q}\right)$ is the intraband contribution
which involves only states with free carriers and is critical for
doped semiconductors. It is computed using Random Phase Approximation
(RPA),
\begin{align}
\epsilon\super{intra}\left(\vec{q}\right)= & 1-V\super{bare}\left(\vec{q}\right)\sum_{\vec{k}mn}\left(\begin{array}{c}
\frac{f_{\vec{k}-\vec{q},m}-f_{\vec{k}n}}{\epsilon_{\vec{k}-\vec{q},m}-\epsilon_{\vec{k},n}}\times\\
|\left\langle u_{\vec{k}-\vec{q},m}|u_{\vec{k}n}\right\rangle |^{2}
\end{array}\right),
\end{align}

where the sum runs over only states having free carriers, e.g., for
a n-doped semiconductor, $m$ and $n$ are conduction band indices.
In the above formula, $f$ is time-dependent non-equilibrium
occupation instead of the equilibrium one $f^{\mathrm{eq}}$. Therefore,
if pump is activated or optical field $\vec{A}_{0}\left(t\right)$ of
the pump pulse is not negligible, $\epsilon\super{intra}\left(\vec{q}\right)$
will be updated in every time step, as $f$ will differ from $f^{\mathrm{eq}}$
and the magnitude of difference depends on the excitation density. $V\super{bare}\left(\vec{q}\right)=e^{2}/\left(V\sub{cell}\varepsilon_{0}|q|^{2}\right)$
is the bare Coulomb potential with $V\sub{cell}$ the unit cell volume
and $\varepsilon_{0}$ vacuum permittivity. $u_{\vec{k}n}$ is the
periodic part of the Bloch wave function.

We then have the matrix elements in reciprocal space,
\begin{align}
g_{13}^{i}= & ZV\super{scr}\left(\vec{q}_{13}\right)\left\langle u_{1}|u_{3}\right\rangle ,\\
g_{1234}\super{e\text{-}e}= & V\super{scr}\left(\vec{q}_{13}\right)\delta_{\vec{k}_{1}+\vec{k}_{2},\vec{k}_{3}+\vec{k}_{4}}\left\langle u_{1}|u_{3}\right\rangle \left\langle u_{2}|u_{4}\right\rangle ,\\
V\super{scr}\left(\vec{q}_{13}\right)= & V\super{bare}\left(\vec{q}_{13}\right)/\epsilon\left(\vec{q}_{13}\right),\label{eq:scr}
\end{align}

where $V\super{scr}\left(\vec{q}\right)$ is the screened Coulomb
potential and $\vec{q}_{13}=\vec{k}_{1}-\vec{k}_{3}$. $\delta_{\vec{k}_{1}+\vec{k}_{2},\vec{k}_{3}+\vec{k}_{4}}$
is Kronecker delta function and means $\vec{k}_{1}+\vec{k}_{2}=\vec{k}_{3}+\vec{k}_{4}$.
$\left\langle u_{1}|u_{3}\right\rangle $ is the overlap matrix element
between two periodic parts of the Bloch wave functions.

\section*{Appendix B: Carrier scattering rate and Im$\Sigma$ from the density-matrix
approach}

At the semiclassical limit, density matrix $\rho$ is replaced by
(non-equilibrium) occupation $f$, then the scattering
term originally with a full quantum description in Eq.~\ref{eq:scattering}
required by DM dynamics becomes:
\begin{align}
\frac{df_{1}}{dt}|_{c}= & \mathop{\sum_{2\neq1}}\left[\left(1-f_{1}\right)P_{11,22}^{c}f_{2}-\left(1-f_{2}\right)P_{22,11}^{c}f_{1}\right],\label{Eq.f1}
\end{align}

using the facts that $P_{11,22}$ is real and ``2=1'' term is zero.
``c'' represent a scattering channel. Note that the weights of k
points must be considered when doing sum over k points.

Suppose $f$ is perturbed from its equilibrium value by $\delta f$,
i.e., $f=f\super{eq}+\delta f$, then insert $f$ after perturbation
into Eq.~\ref{Eq.f1} and linearize it,
\begin{align}
\frac{df_{1}}{dt}|_{c}= & -\mathop{\sum_{2\neq1}}\left[P_{11,22}^{c}f_{2}^{\mathrm{eq}}+\left(1-f_{2}^{\mathrm{eq}}\right)P_{22,11}^{c}\right]\delta f_{1},
\end{align}

using the fact that $\delta P_{11,22}$ is always zero, even for the
e-e scattering.

Define carrier relaxation time of state ``1'' $\tau_{p,1}^{c}$
by $\frac{df_{1}}{dt}|_{c}=-\frac{\delta f_{1}}{\tau_{p,1}^{c}}$,
we have
\begin{align}
\frac{1}{\tau_{p,1}^{c}}= & \mathop{\sum_{2\neq1}}\left[P_{11,22}^{c}f_{2}^{\mathrm{eq}}+\left(1-f_{2}^{\mathrm{eq}}\right)P_{22,11}^{c}\right].\label{eq:carrier_lifetime}
\end{align}

The linewidth or the imaginary part of the self-energy for the scattering
channel $c$ is related to the carrier relaxation time by $\mathrm{Im}\Sigma_{1}^{c}=\hbar/\left(2\tau_{p,1}^{c}\right)$.

Using Eq.~\ref{eq:carrier_lifetime}, we have calculated the e-ph
scattering rates and they are in good agreement with previous theoretical
results \citep{zhou2016ab}. For e-ph scattering, Eq.~\ref{eq:carrier_lifetime}
will reproduce the imaginary part of the well-known Fan-Migdal self-energy\citep{giustino2017electron}.

For e-i scattering, we have 
\begin{align}
\frac{1}{\tau_{p,1}\super{e\text{-}i}}= & \frac{2\pi}{\hbar}n_{i}V\sub{cell}\sum_{2}|g_{12}^{i}|^{2}\delta_{\sigma}^{G}\left(\epsilon_{1}-\epsilon_{2}\right).\label{tau-e-i}
\end{align}

The above equation (Eq.~\ref{tau-e-i}) is consistent with Ref. \citenum{jacoboni2010theory}.

\begin{figure}
\includegraphics[scale=0.38]{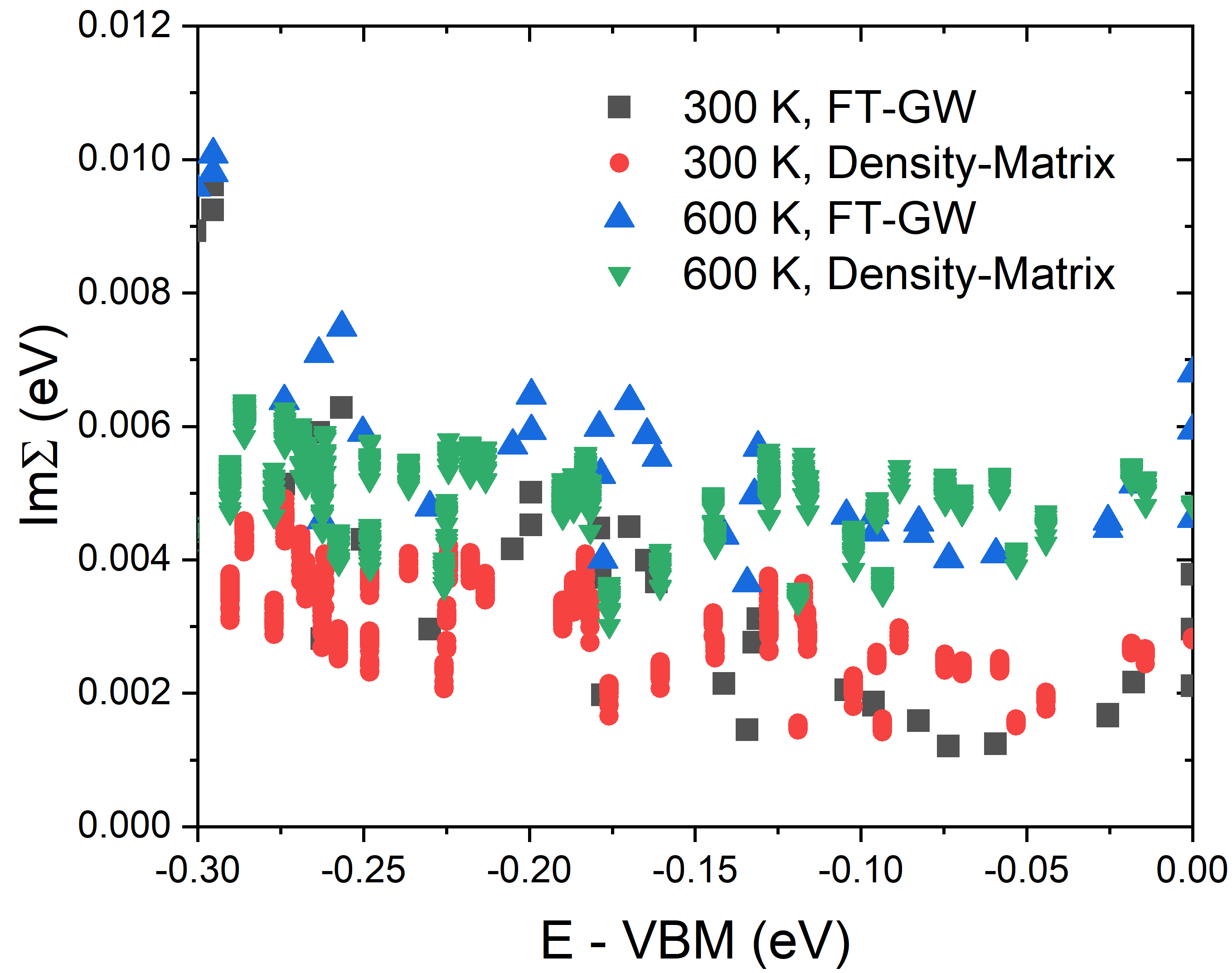} \caption{$\mathrm{Im\Sigma}$ due to e-e scattering of valence electrons of
p-type silicon computed by Eq.~\ref{eq:ee_lifetime} (Density-Matrix)
compared with those calculated by the finite-temperature GW method
(FT-GW) \citep{benedict2002quasiparticle}. $\mu$ is set to 0.05
eV lower than Valence Band Maximum (VBM). For simplicity, SOC is not
considered in this test.\label{fig:ImSigma_benchmark}}
\end{figure}

For e-e scattering, neglecting the exchange contribution, which is
a commonly-used approximation\cite{rossi2002theory,benedict2002quasiparticle},
\begin{align}
\frac{1}{\tau_{p,1}\super{e\text{-}e}}= & \frac{2\pi}{\hbar}\sum_{2\neq1,34}|A_{1324}|^{2}\left[\begin{array}{c}
f_{2}^{\mathrm{eq}}f_{4}^{\mathrm{eq}}\left(1-f_{3}^{\mathrm{eq}}\right)+\\
\left(1-f_{2}^{\mathrm{eq}}\right)f_{3}^{\mathrm{eq}}\left(1-f_{4}^{\mathrm{eq}}\right)
\end{array}\right].\label{eq:ee_lifetime}
\end{align}
To verify our implementation of e-e scattering term, we have calculated
$\mathrm{Im\Sigma}$ due to e-e scattering of valence electrons of
p-type silicon based on the above equation and compare it with those
calculated by the finite-temperature GW method from first-principles,
\citep{benedict2002quasiparticle} as implemented in JDFTx.\citep{sundararaman2017jdftx}
The JDFTx implementation, in turn, has been benchmarked to reproduce
the expected dependence with temperature and carrier energy, Im$\Sigma\super{e\text{-}e}\propto(\varepsilon-\varepsilon_{F})^{2}+(\pi k_{B}T)^{2}$,
as expected for metals.\citep{TAparameters}

From Fig.~\ref{fig:ImSigma_benchmark}, we can see the results by
two methods agree well for the energy range close to the Fermi level
which is relevant to e-e scatterings due to energy conservation. 
This verifies our implementation of e-e scattering part.

\section*{Appendix C: The effects of $\omega\sub{pump}$ and pump fluence on
spin relaxation of GaAs at 300 K}

\begin{figure}
\includegraphics[scale=0.38]{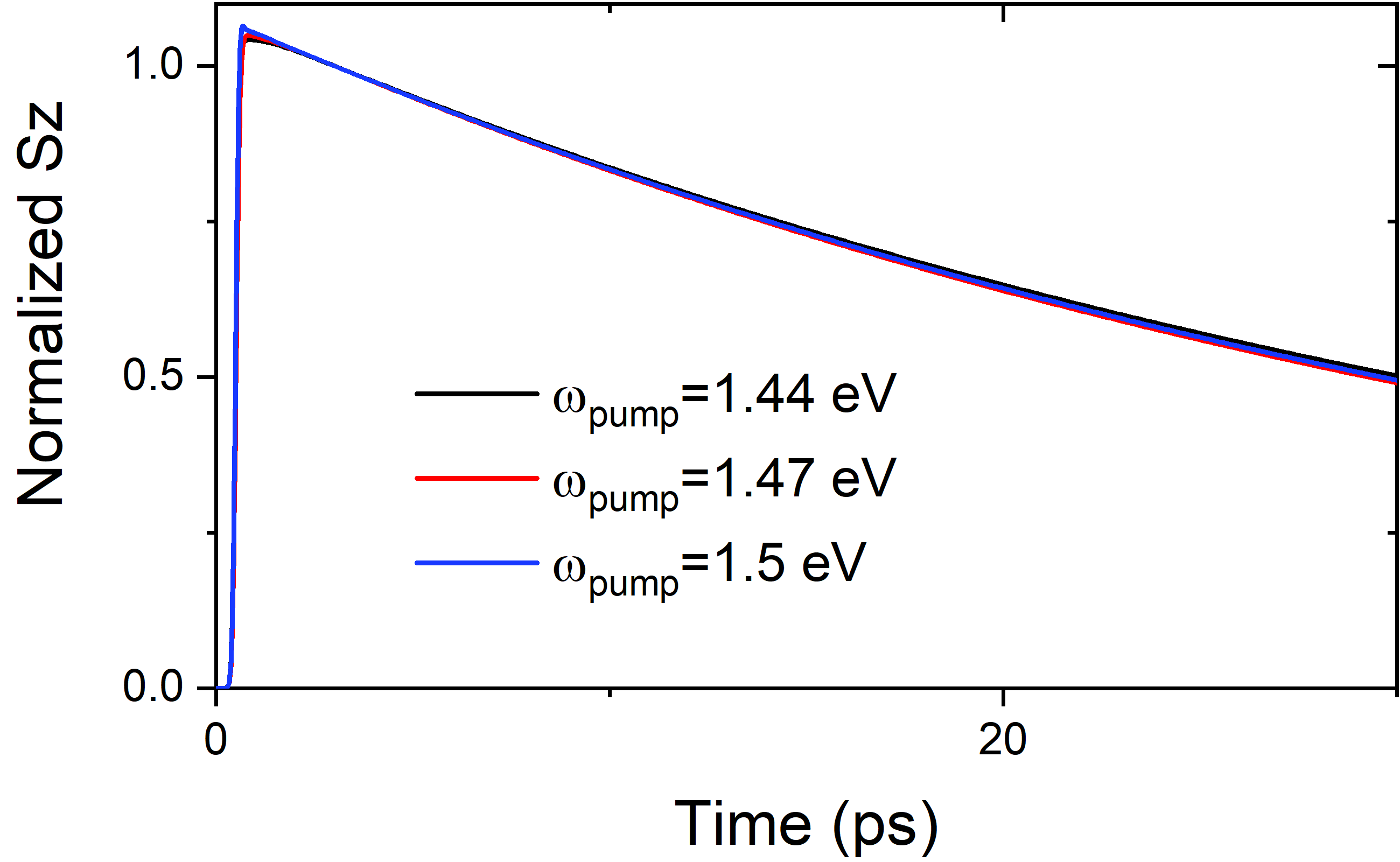}

\caption{$S_{z}\left(t\right)$ of $n$-GaAs with $n_{i}=2\times10^{16}$ cm$^{-3}$
at 300 K with different pump pulse energies ($\omega\sub{pump}$)
varying with several $k_{B}T$. \label{fig:pumpE}}
\end{figure}

In Fig.~\ref{fig:pumpE}, we study the $S_{z}$ relaxation dependence
on pump-pulse energy changes with several $k_{B}T$. We can see that
variation of $\omega\sub{pump}$ has very weak effects on spin dynamics
of $n$-GaAs at 300 K.

\begin{figure}
\includegraphics[scale=0.24]{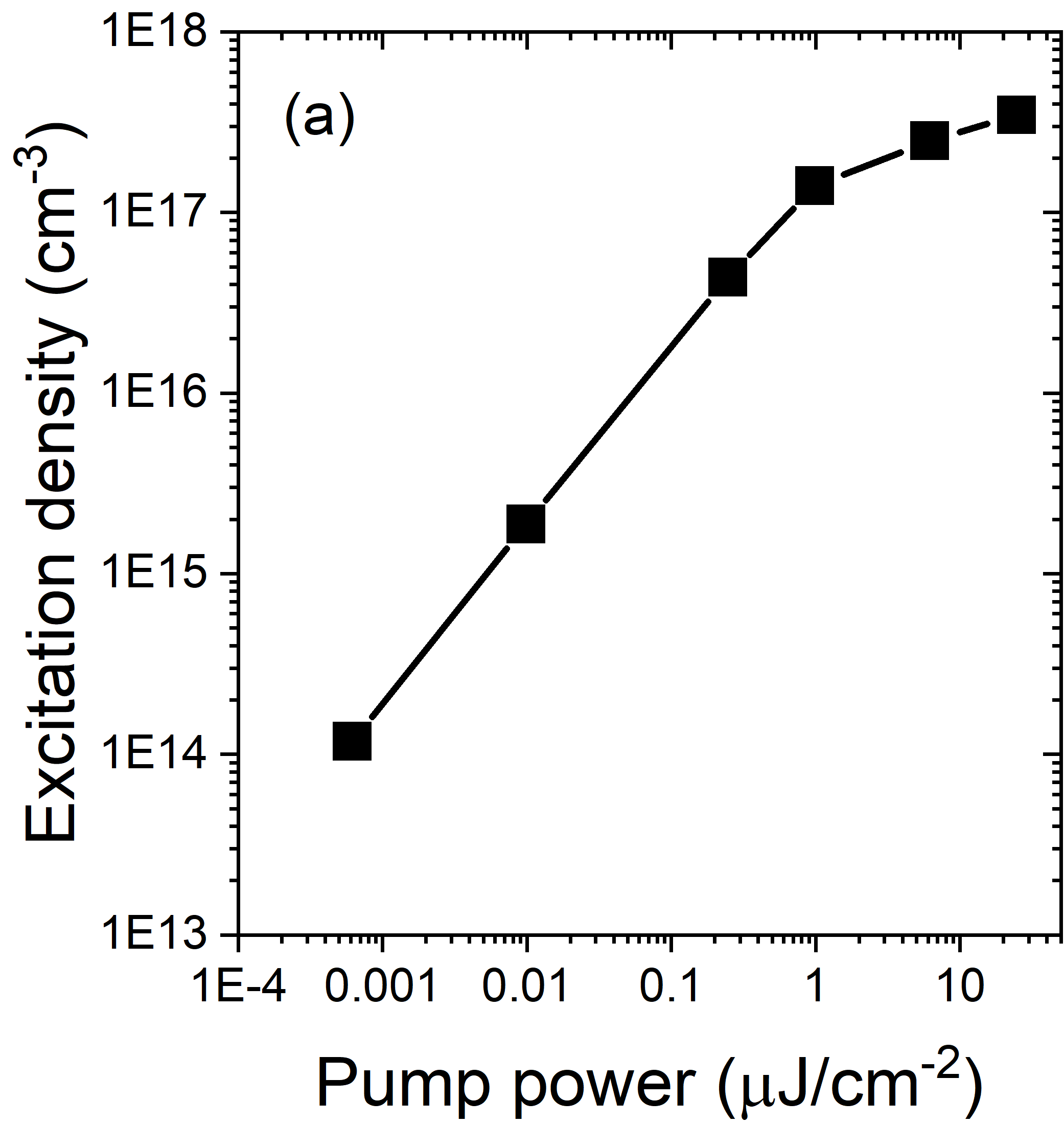} \includegraphics[scale=0.24]{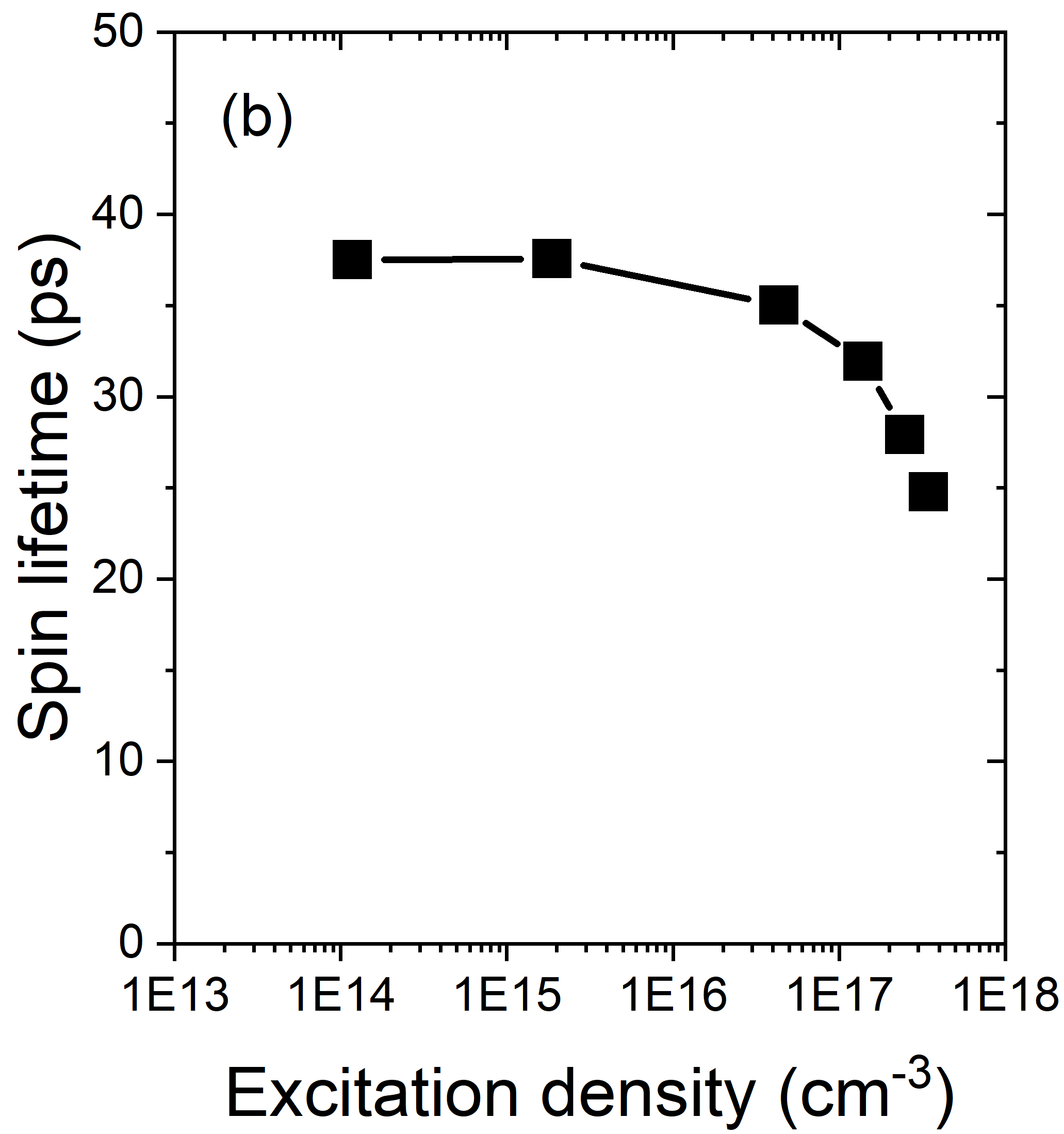}
\caption{(a) The excitation density as a function of the pump fluence (left panel)
and (b) the spin lifetime as a function of the excitation density
generated by a circularly polarized pump pulse for
$n$-GaAs with $n_{i}=10^{14}$ cm$^{-3}$ at 300 K. $\omega\sub{pump}$=1.47
eV.\label{fig:pumpfluence}}
\end{figure}

In Fig.~\ref{fig:pumpfluence}a, we study the effects of the pump fluence
$I\sub{pump}$ on spin relaxation. Firstly, we can see that in the
low pump fluence region or when $I\sub{pump}<1$ $\mu$J/cm$^{-2}$,
the excitation density increases linearly with $I\sub{pump}$ but
when $I\sub{pump}>1$ $\mu$J/cm$^{-2}$, the excitation density increases
slower. This is because in high fluence cases, during the excitation
by a pump pulse, a significant amount of conduction states have been
already filled, which reduce the probability of the transitions from
valence bands to conduction bands. From Fig.~\ref{fig:pumpfluence}b,
we find that spin lifetime of $n$-GaAs decreases with the excitation
density. This dependence may be explained based on the empirical DP
relation\citep{vzutic2004spintronics} $\tau_{s,i}\sim\tau_{s,i}^{DP}=1/\left[\overline{\tau}_{p}\cdot\left(\left\langle \vec{\Omega}^{2}\right\rangle -\left\langle \Omega_{i}^{2}\right\rangle \right)\right]$
as we discussed in Sec.~\ref{subsec:temperature}. At 300 K, generally
$\left\langle \vec{\Omega}^{2}\right\rangle -\left\langle \Omega_{i}^{2}\right\rangle $
will increase with increasing free carrier density (through an increase
of excitation density here), similar to what we find at 30 K shown
in Fig. \ref{fig:doping}c. On the other hand, $\overline{\tau}_{p}$
due to the electron-phonon scattering, which dominates carrier relaxation
at 300 K, is less sensitive to the variation of excitation density.
Therefore, it is the increase of $\left\langle \vec{\Omega}^{2}\right\rangle -\left\langle \Omega_{i}^{2}\right\rangle $
causing the decrease of spin lifetime when increasing excitation density.

\section*{Appendix D: Layer-resolved quantities of bilayer WSe$_{2}$}

For a bilayer with one layer above $z=0$ and another
below, the operator $\hat{l}^{\mathrm{top}}$ projecting any local
quantity $A(\vec{r})$ to the top layer can be defined through the
relation
\begin{align}
\hat{l}^{\mathrm{top}}A(\vec{r})= & H\left(z\right)A(\vec{r}),
\end{align}

where $z$ is the third component of $\vec{r}$ and
$H\left(z\right)$ is Heaviside step function
\begin{align}
H\left(z\right)= & \{\begin{array}{c}
1,z>0\\
0,z\leq0
\end{array}.
\end{align}

Therefore, top layer hole occupation $f_{h}^{\mathrm{top}}$ is
\begin{align}
f_{h}^{\mathrm{top}}= & \mathrm{Tr}\left(\hat{l}^{\mathrm{top}}\left(1-f\right)\right)\\
= & \sum_{kn}l_{k,nn}^{\mathrm{top}}\left(1-f_{kn}\right),
\end{align}

where $f$ is occupation and $l_{k,mn}^{\mathrm{top}}=\left\langle km\right|\hat{l}^{\mathrm{top}}\left|kn\right\rangle =\left\langle km\right|H\left(z\right)\left|kn\right\rangle $.
$\mathrm{Tr}$ means taking trace. $k$ is k-point index. $n$ and
$m$ are band indices.

Moreover, top layer spin $s_{z}^{\mathrm{top}}$
can be defined using the operator $\hat{l}^{\mathrm{top}}\hat{s}_{z}$,
\begin{align}
s_{z}^{\mathrm{top}}= & \mathrm{Tr}\left(\hat{l}^{\mathrm{top}}\hat{s}_{z}\hat{\rho}\right)\\
= & \sum_{k,lmn}l_{k,lm}^{\mathrm{top}}s_{z,k,mn}\rho_{k,nl}.
\end{align}

Note that the commutator $\left[l^{\mathrm{top}},s_{z}\right]$
is found numerically close to zero for bilayer WSe$_2$.
This indicates that we can safely define $s_{z}^{\mathrm{top}}$ 
as an observable using the above equations.


%

\end{document}